\documentclass[prl,twocolumn,showpacs,10pt,floatfix,superscriptaddress, nofootinbib,reprint]{revtex4-1}
\usepackage{amsmath,amssymb,graphicx,bm,color,float,ulem}
\usepackage{amsfonts,amsthm}
\usepackage{epstopdf}
\usepackage{xcolor}

\usepackage{pdfpages}
\makeatletter
\AtBeginDocument{\let\LS@rot\@undefined}
\makeatother

\begin{document}

\title{Isolated, attosecond spatio-temporal vortices 
from high-order harmonics with non-scaling topological charge} 
\title{Isolated attosecond spatio-temporal optical vortices: Interplay between the topological charge and orbital angular momentum scaling in high harmonic generation}

\author{Rodrigo Mart\'{\i}n-Hernández}
\email{rodrigomh@usal.es}
\address{Grupo de Investigación en Aplicaciones del Láser y Fotónica, Departamento de F\'{i}sica Aplicada, Universidad de Salamanca, 37008, Salamanca, Spain} 
\address{Unidad de Excelencia en Luz y Materia Estructuradas (LUMES), Universidad de Salamanca, Salamanca, Spain}
\author{Luis Plaja}
\address{Grupo de Investigación en Aplicaciones del Láser y Fotónica, Departamento de F\'{i}sica Aplicada, Universidad de Salamanca, 37008, Salamanca, Spain} 
\address{Unidad de Excelencia en Luz y Materia Estructuradas (LUMES), Universidad de Salamanca, Salamanca, Spain}
\author{Carlos Hernández-Garc\'{\i}a}
\address{Grupo de Investigación en Aplicaciones del Láser y Fotónica, Departamento de F\'{i}sica Aplicada, Universidad de Salamanca, 37008, Salamanca, Spain} 
\address{Unidad de Excelencia en Luz y Materia Estructuradas (LUMES), Universidad de Salamanca, Salamanca, Spain}
\author{Miguel A. Porras}
\email{miguelangel.porras@upm.es}
\address{Grupo de Sistemas Complejos, ETSIME, Universidad Politécnica de Madrid, Rios Rosas 21, 28003 Madrid, Spain
}

\begin{abstract}
The propagation properties and the nature of the transverse orbital angular momentum (t-OAM) of spatiotemporal optical vortices (STOVs) open new scenarios in high-harmonic generation (HHG), where the richness of the topological charge and OAM up-conversion are exposed. Through advanced numerical simulations, we demonstrate that HHG driven by spatio-spectral optical vortices produces far-field, extreme-ultraviolet STOV harmonics with non-scaling topological charge, i.e., with the same topological charge. This allows for the generation of attosecond STOVs, in contrast to previous works of HHG driven by STOVs, where the topological charge scales with the harmonic order. Our findings evidence that the scaling of the topological charge in HHG driven by spatio-temoral topological fields is not generally connected to that of the up-converted OAM. The up-converted intrinsic OAM does scale with generality with the harmonic order in HHG, albeit this scaling does not necessarily imply its conservation.

\end{abstract}


\maketitle

Spatiotemporal optical vortices (STOVs) are fundamental light fields endowed with a spatiotemporal topology. They carry a phase line singularity along a direction transverse to the propagation direction, around which the spatiotemporal phase twists. Prototypical STOVs have a spatiotemporal intensity ring surrounding the singularity, and carry a transverse orbital angular momentum (t-OAM) proportional to the vortex topological charge ($\ell$) \cite{Bliokh2012,Bliokh2021,Bliokh2023,porras2023transverse,porras2024clarification}. After their first experimental observation in a nonlinear process \cite{Jhajj2016}, STOVs have been generated in the infrared (IR) using a variety of adapted $4f$ pulse shapers \cite{Hancock19,chong2020,Huan2024SciAdv}.
Novel proposals exploit photonic crystal slabs \cite{wang2021}, metasurfaces \cite{huang2023spatiotemporal} and nanogratings \cite{huo2024observation}. The generated STOVs have found application as information carriers \cite{Huan2024SciAdv}, in electron acceleration \cite{Sun2024}, or as building blocks of more complex topological spatiotemporal fields \cite{wan2022scalar}.

The generation of STOVs at higher photon energies, up to the extreme-ultraviolet (EUV) or soft x-rays, is challenging as standard refractive optical elements become nearly transparent. This limitation can be circumvented using non-linear optics. For instance, near-IR STOVs can be up-converted into the visible/ultraviolet using second \cite{gui2021,hancock2021}, third harmonic generation \cite{Wang23third}, and frequency up-conversion \cite{Gao23}. Non-perturbative high harmonic generation (HHG) provides an appealing alternative not only to extend their photon energies, but to synthesize them into attosecond pulses. HHG is a highly nonlinear process, in which an intense IR driving field is focused into a gas target, resulting in the generation of high-order harmonics that can reach the EUV or soft-X rays \cite{McPherson1987, Ferray1988, Popmintchev2012}. The high-order harmonics are inherently phase-locked, which enables their synthesization into attosecond pulses \cite{Farkas1992, Antoine1996, Christov1997, Paul2001, Hentschel2001}. In addition, thanks to its extreme coherence, HHG allows a controlled up-conversion of the driving field properties into the harmonics. In particular, if driven by a longitudinal vortex beam with topological charge $\ell$, the topological charge of the $q^{\rm th}$ order harmonic scales as $\ell_q=q\ell$ \cite{Hernandez-Garcia2013, Gariepy2014, Geneaux2016, Pandey2022} (unless multiple driving fields involving different spin and $\ell$ are used \cite{Turpin2017, Gauthier2017, Kong2017, dorney_2019_SAMOAMHHG, Pisanty2019, delasHeras_2022, luttmann2023nonlinear, de2024attosecond}). Recently, HHG driven by STOVs has been theoretically proposed \cite{fang2021}, and experimentally realized \cite{Martin-Hernandez2025spatiotemporal}, demonstrating the generation of EUV harmonic STOVs whose topological charge also follows the $\ell_q=q\ell$ scaling law. This scaling precludes  the generation of harmonic STOVs with the same topological charge, and thus their synthesization into attosecond STOVs.

In this Letter, we demonstrate that high-order harmonic STOVs with non-scaling $\ell_q$, i.e., with the same topological charge, can be generated by driving HHG with a spatio-spectral optical vortex (SSOV), the spectral counterpart of a STOV.
As a main result, this non-scaling property enables the synthesization of the high-order harmonics into an attosecond STOV. Advanced numerical simulations that properly describe both the quantum-matter and the macroscopic aspects of the HHG process support our theoretical predictions, showing that an attosecond STOV is produced within a train of attosecond pulses. We further show that by imprinting spatial chirp to the SSOV driving field, i.e., implementing the lighthouse effect \cite{vincenti2012attosecond,  wheeler2012attosecond, kim2013photonic, hammond2016attosecond} in the transverse spatial coordinate, the attosecond STOV can be spatially isolated from the train.

In addition, we unveil the physics that enables the non-scaling $\ell_q$ in SSOV-driven HHG. In photon up-conversion processes with longitudinal vortices of Laguerre-Gauss (LG) type, the scaling law $\ell_q = q\ell_0$ is identified with conservation of OAM, since $\ell$ and OAM per photon ---a purely intrinsic, origin independent OAM \cite{ONeil}--- are proportional \cite{chong2020,Hernandez-Garcia2013, Gariepy2014, Geneaux2016, Pandey2022}. 
Unlike LG vortices, first, STOV/SSOVs may have an intrinsic t-OAM, an OAM of inner wave rotation ---the only part of the OAM that may be regarded as a property of photons--- differing from the total OAM about a fixed point or axis \cite{Bliokh2012,Bliokh2021,Bliokh2023,porras2023transverse,porras2024clarification}. 
Second, $\ell$ and intrinsic OAM per photon are proportional only in particular situations \cite{porras2023propagation,porras2023transverse}. Here we evaluate the intrinsic, longitudinal OAM (l-OAM) and t-OAM in HHG driven by standard LG vortices, STOVs and SSOVs. We find the intrinsic OAM per photon as the physical magnitude that scales with the harmonic order with generality, regardless of the scaling of the topological charge. While the intrinsic l-OAM is conserved ---the intrinsic l-OAM per photon of the $q^{\rm th}$ harmonic order is $\hbar \ell_q = q\hbar\ell$, as expected from the existence of individual photons with quanta of l-OAM,--- the intrinsic t-OAM in HHG driven by STOVs and SSOVs is not generally conserved, but may take continuous, arbitrary values. These are fundamental observations that set out a more comprehensive basis
for further research on nonlinear processes driven by topological spatiotemporal light fields.



\begin{figure}
\includegraphics[width=0.5\textwidth]{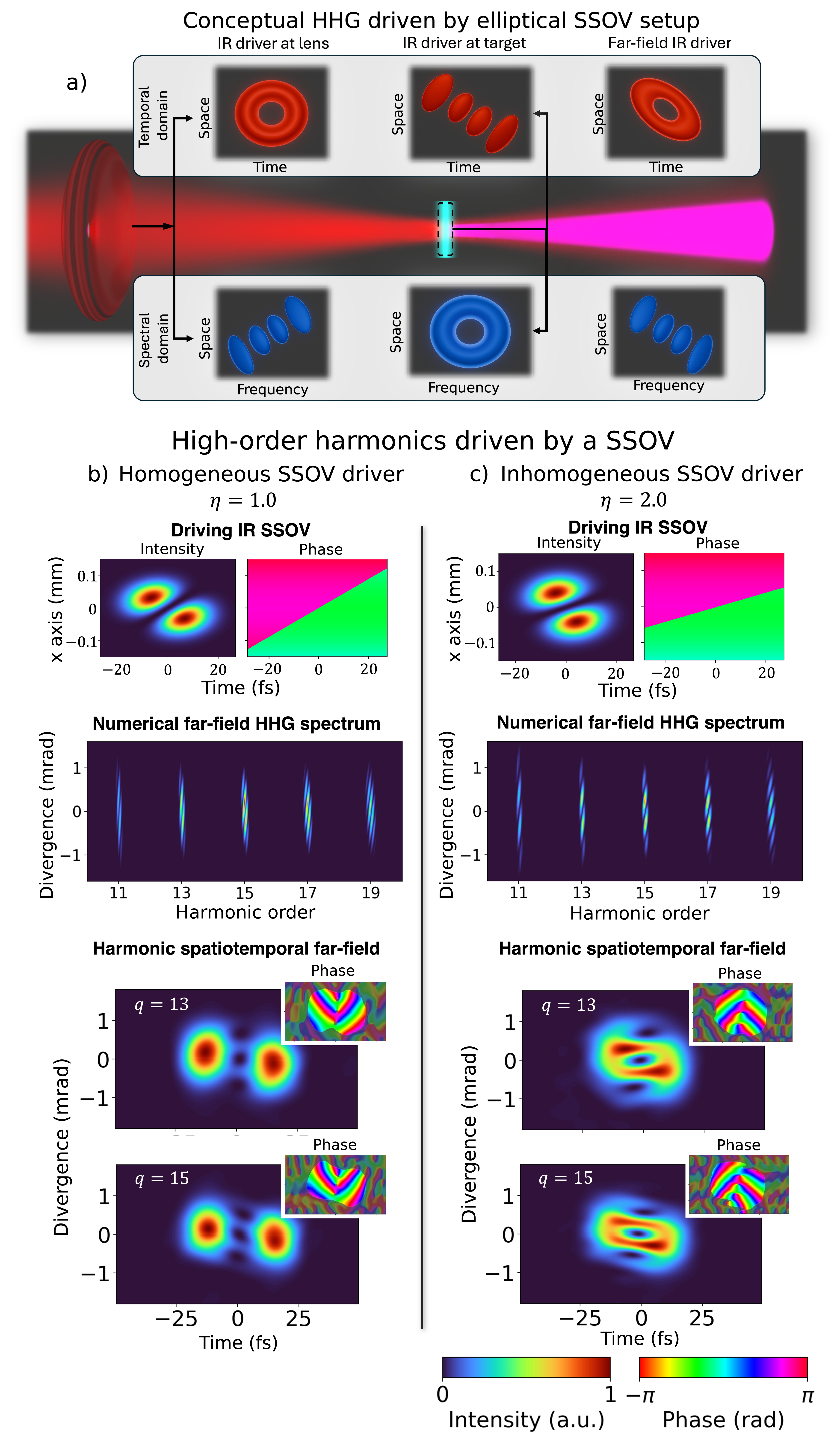}
\caption{(a) Conceptual depiction of the intensity of a focused IR STOV for HHG driven by a SSOV. The top panels in red illustrate the spatiotemporal field evolution from the focusing system, to the gas-jet and to the far-field. The bottom panels in blue are the spatio-spectral view, with the SSOV at the jet. (b) IR spatiotemporal driving field with $\eta=1$ at the gas-jet, the far-field, high-order, angularly resolved harmonic spectra, and the spatiotemporal intensity and phase of the 13$^{\rm th}$ and 15$^{\rm th}$ harmonics, obtained from the numerical simulations. (c) The same but for $\eta=2$. The simulation parameters for (b) and (c) are provided in the text.
} 
\label{fig1}
\end{figure}

An elliptical STOV with the phase line singularity perpendicular to the $x$-$t$ plane [Fig. \ref{fig1}(a), left-top panel] is described by
\begin{equation}\label{eq:stov}
E(x,t)=E_0\left(\frac{t}{\eta t_0}\mp i \frac{x}{x_0}\right)^{|\ell|} e^{-\frac{x^2}{x_0^2}}e^{-\frac{t^2}{t_0^2}} e^{-i\omega_0 t},
\end{equation}
where $\omega_0$ is the carrier frequency, $x_0$ the Gaussian width, $t_0$ the Gaussian duration, and the $\mp$ sign stands for positive and negative $\ell$. The inhomogeneity parameter $\eta$ distorts a STOV from its perfect elliptical shape ($\eta=1$). A Gaussian profile $e^{-y^2/y_0^2}$ along $y$, with $y_0=x_0$, is assumed, which remains factorized upon propagation, and hence is not written. The focused field at any distance $z$ beyond a lens of focal length $f$ is \cite{Porras_nano_2025}
\begin{align}
\begin{split}\label{eq:focusing}
&E(x,t',z) =E_0\sqrt{\frac{q_1}{q_{2}}} e^{\frac{ik_0x^2}{2q_{2}}} e^{-\frac{t^{\prime 2}}{t_0^2}}\left(\frac{iq z}{q_{2} z_{R}}\right)^{\frac{|\ell|}{2}}  \\ 
&\times\frac{1}{2^{|\ell|}} H_{|\ell|}\left[\sqrt{\frac{q_{2} z_{R}}{iq_1 z}}\left(\frac{t'}{\eta t_0}\mp i\frac{x}{x_0}\frac{q_1}{q_{2}}\right)\right] e^{-i\omega_0 t'},
\end{split}
\end{align}
where $k_0=\omega_0/c$ is the propagation constant, $t'=t-z/c$ is the local time, $q_{2}=q_1+z$, $1/q_1=(-1/f+1/q_{0})$, and $1/q_{0}= i/z_{R}$ are Gaussian complex beam parameters at $z$, after and before the lens, $z_{R}=k_0x_0^2/2$, and $H_{|\ell|}(\cdot)$ is the Hermite polynomial of order $|\ell|$. According to Eq. (\ref{eq:focusing}), the STOV degenerates at the focal plane ($z=f$) into a spatiotemporal tilted-Hermite lobulated (ST-THL) field with $|\ell|$ tilted lines of zero amplitude where the phase experience $\pi$ jumps, between $|\ell|+1$ tilted lobes of increasing intensity outwards [Fig. \ref{fig1}(a), middle-top panel]. A quasi-elliptical STOV is recovered at the far field [Fig. \ref{fig1}(a), right-top panel], but of opposite $\ell$ \cite{porras2023transverse,Porras_nano_2025}. 

Owing to the duality between the transverse spatial and temporal spectra of STOVs \cite{Porras_nano_2025}, these STOV/ST-THL/STOV transmutations upon propagation are seen the opposite in spatiospectral domain [Fig. \ref{fig1}(a), bottom panels]. The spatiospectral tilted-Hermite lobulated (SS-THL) focuses to a SSOV, and then diverges to a SS-THL spatiospectrum.

We have performed advanced numerical HHG simulations driven by an IR ST-THL field, i.e., a SSOV, in a gas jet placed at the focus ($z=f$), as depicted in Fig. \ref{fig1}(a). The simulations consider both the microscopic quantum dynamics of the individual emitters and the macroscopic phase matching of the whole target. To this end, the individual atomic dipole acceleration is calculated through the full-quantum strong field approximation, which, combined with the Maxwell far-field propagator, yields the harmonics' collective macroscopic far-field emission \cite{Hernandez-Garcia2010}. This method has been successfully compared against several previous HHG experiments driven by structured beams \cite{Pandey2022, HernandezGarcia2017, delasHeras_2022, dorney_2019_SAMOAMHHG, de2024attosecond, Rego2019torque}, including HHG driven directly by STOVs (or SS-THL) \cite{Martin-Hernandez2025spatiotemporal}. In the HHG numerical simulations, we consider a $\sin^2(\pi t/t_{\rm max})$ temporal envelope, where $t_{\rm max} = 32$ optical cycles (equivalent to $t_0\approx 26.11$ fs of a Gaussian envelope). The driving field, with a peak intensity of $1.6\times10^{14}\:{\rm W/cm^2}$ and a central wavelength of $800\:{\rm nm}$, is focused into an infinitesimally thin atomic hydrogen gas-jet ---an approach that has been validated against experimental results \cite{Martin-Hernandez2025spatiotemporal}. Atomic hydrogen is used for computational simplicity, but the results presented here are universal to any noble gas. 

Fig. \ref{fig1}(b) shows the results of HHG driven by a ST-THL field with a perfectly elliptical SSOV ($\eta=1$) formed upon focusing a STOV with  $x_0=1\:\rm mm$, $t_0 = 26.11$ fs, and $\ell=1$ at a lens of focal length $f=250\:\rm mm$. In the top panel we depict the driving intensity and phase spatiotemporal profiles. In the central panel the far-field spatiospectral intensity profile of high-order harmonics is shown. Each harmonic distribution is split into several tilted lobes, which, according to Fig. \ref{fig1}(a), suggests the presence of a STOV in the temporal domain. The bottom panel shows the far-field spatiotemporal intensity and phase distributions of two harmonics (13th and 15th). Remarkably, the phase distributions show several single-charged phase singularities aligned at $t=0$ along the divergence axis, indicating that 
the $\ell_q$ of the spatiotemporal profile does not scale with the harmonic order. This results contrasts to LG-driven and STOV-driven HHG \cite{Hernandez-Garcia2013,Martin-Hernandez2025spatiotemporal}, where $\ell_q=q\ell$. Indeed,  $\ell_q=-\ell$, where the {sign reversal reflects} the same reversal of the driving field when propagating {from the lens} to the far field \cite{porras2023transverse,Porras_nano_2025}. 

The spatio-temporal harmonic intensity distributions in Fig. \ref{fig1}(b)
show two predominant lobes, but they can be tuned to a more uniform distribution around the central phase singularity trough the inhomogeneity parameter $\eta$. Fig. \ref{fig1} (c) shows the HHG results when the driving STOV at the lens is distorted with $\eta=2$. The resulting far-field harmonic STOVs still exhibit the single-charged singularities, but the harmonic intensity is distributed around the central phase singularity more uniformly, resulting in EUV STOVs with unit {$|\ell|$.}

\begin{figure}[!b]
\includegraphics[width=0.45\textwidth]{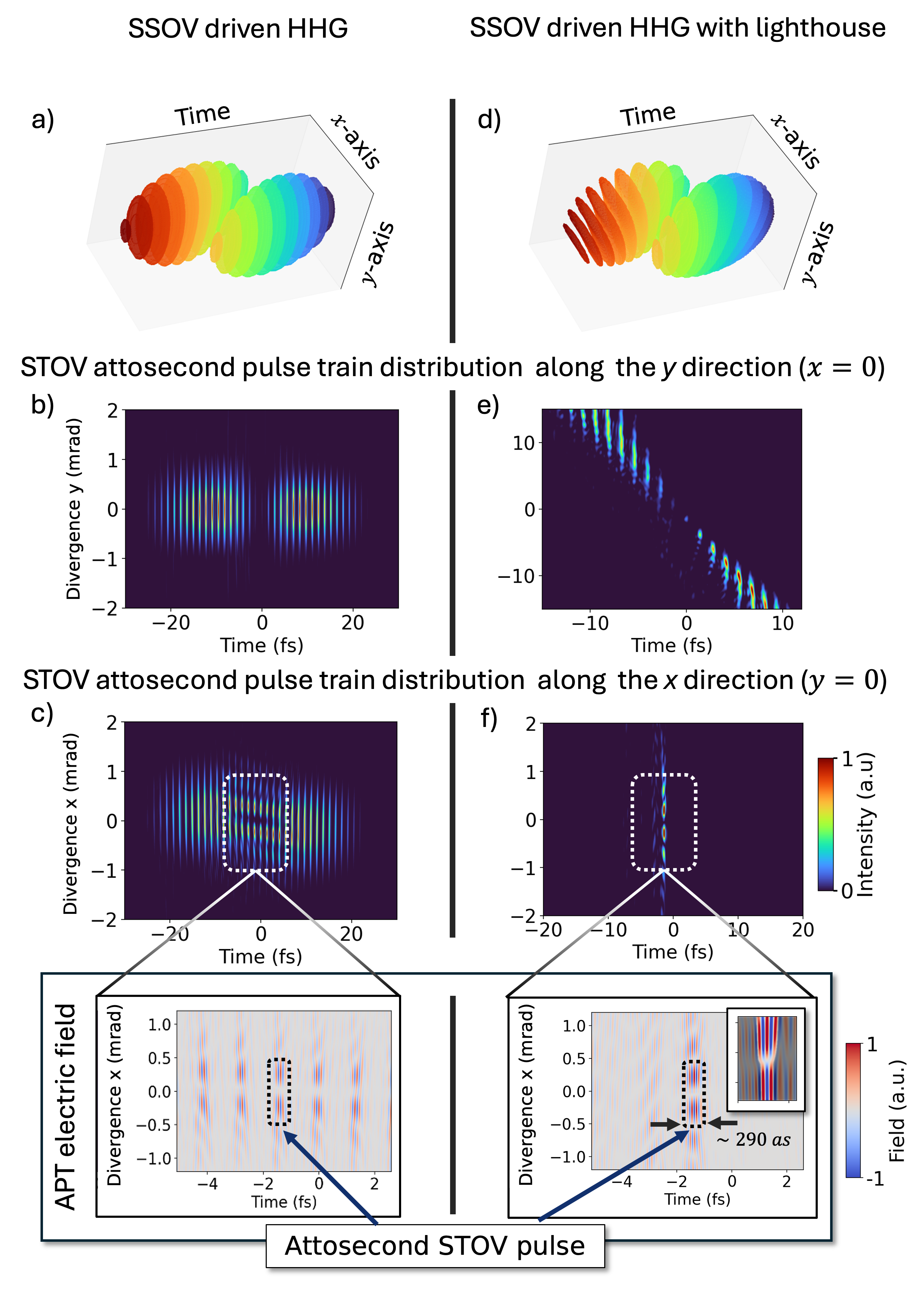}
\caption{Generation of attosecond EUV-STOV pulses from HHG driven by SSOV fields ($\eta=2$). The driving field parameters are the same as in Fig. \ref{fig1}. (a) Driving SSOV wavefront isosurface at the gas-jet. (b) and (c) represent the attosecond pulse train obtained by the superposition of the harmonics order from $13^{\rm th}$ to $19^{\rm th}$. Panel (b) corresponds with the attosecond pulse train distribution along $y-t$ plane,  with $x=0$, while (c) shows the profile in the $x-t$ plane, with $y=0$. Note the spatiotemporal singularity at the center of the pulse train. (d)-(f) illustrate the SSOV-driven HHG results with the additional lighthouse effect --- rotating wavefront as depicted in panel (d)--- along the $y$-direction with $\beta=1.56\times10^{-2}\:{\rm \mu m^{-1} fs^{-1}}$. In this case each pulse of the train is emitted at different angles along the $y$-axis, panel (e). The attosecond pulse carrying the spatiotemporal singularity is the only one emitted on axis, shown in panel (f). The zoomed inset illustrate the real field of the isolated attosecond STOV pulse with $\sim 290\:\rm as$ full-width half-maximum pulse duration.}
\label{fig2}
\end{figure}

The interpretation as a disorted STOV of unit {$|\ell|$} also follows from the simplest HHG model. The $q^\textrm{th}$ order harmonic field can be expressed as $E_q\simeq|E|^{\rm q_{eff}} e^{iq\,{\rm arg}(E)}e^{i\phi_{\rm int}}$, where $E$ is the driving field amplitude, $\rm q_{eff}$ is the non-perturbative scaling parameter (typically ${\rm q_{eff}}\sim 3.5$ \cite{Huillier1992}), and $\phi_{\rm int}$ is the intrinsic dipole harmonic phase \cite{lewenstein1995phase}, which plays here a secondary role {(see Sup. Matt)}. In LG-driven and STOV-driven HHG \cite{Hernandez-Garcia2013,Martin-Hernandez2025spatiotemporal}, the harmonic phase is given by ${\rm arg}(E_q) \simeq ql\varphi-q\omega_0 t'$, where $\varphi$ is an azimuthal spatial or spatiotemporal phase. As a result, the topological charge follows the scaling law $\ell_q=q\ell$. In opposite, in SSOV-driven HHG, the spatiotemporal phase distribution of the ST-THL field, exhibiting a $\pi$-step [Figs. \ref{fig1}(b) and (c) top], is preserved upon up-conversion when multiplied {by $q$}, and, upon propagation to the far field, a harmonic STOV of unit {$|\ell|$} is obtained. However, upon up-conversion, the amplitude profile of the ST-THL field is distorted by the scaling with ${\rm q_{eff}}$. As a consequence, each far-field harmonic corresponds to a distorted STOV of unit {$|\ell|$} with a number of satellite phase singularities, whose spectra exhibit two main tilted lobes with a number of satellite lobes. 
Similar formation of spurious vortices surrounding an imperfect STOV formed from distorted ST-THL fields has been described in \cite{Porras_PRA_2024}.



The generation of harmonic STOVs with same topological charge ($\ell_q=-\ell$) enables their synthesization into an attosecond STOV. Previous proposals and experiments of HHG driven by IR-STOV fields \cite{fang2021, Martin-Hernandez2025spatiotemporal}, where $\ell_q=q\ell$, precluded the generation of attosecond STOVs as their phase distribution strongly varies between the different harmonics. 
Figs. \ref{fig2} (b) and (c) show the spatiotemporal attosecond distribution at planes $x=0$ and $y=0$ respectively, obtained after overlapping from the 13$^\text{th}$ to the 19$^\text{th}$ harmonic orders generated from the inhomogeneous SSOV with $\eta=2.0$ [Fig. \ref{fig1} (c)], 
We observe an attosecond pulse train profile in both $x$- and $y$-directions, with a zero of intensity at the center as a consequence of the spatiotemporal phase singularity. The zoomed inset in Fig. \ref{fig2} (c) shows the real electric field of the attosecond pulse train distribution. Remarkably, the central pulse within the train---and only the central one---exhibits a fork-like dislocation, indicating the presence of an attosecond STOV pulse. 


In order to isolate the attosecond STOV from the train, we propose to imprint an angular chirp in the driving STOV in the $y$-direction---which corresponds to transverse spatial chirp of the driving SSOV at the $y$-direction. Angular chirp is used in HHG to spatially separate attosecond pulses within the train, a technique known as attosecond lightouhse \cite{vincenti2012attosecond, wheeler2012attosecond, kim2013photonic, hammond2016attosecond}.
The resulting rotating wavefront, depicted in Fig. \ref{fig2} (d), is included in Eq. (\ref{eq:focusing}) as $e^{-y^2/y_{0,f}^2 + i\beta yt}$, where $\beta$ accounts for the wavefront rotation velocity and $y_{0,f}=2f/k_0y_0$ is the beam width at focus. Figs. \ref{fig2} (e) and (f) show the resulting attosecond pulse train distribution at planes $x=0$ and $y=0$, respectively. 
Each attosecond pulse is emitted at different angles, as shown in Fig. \ref{fig2} (e), while the attosecond STOV pulse propagates, isolated, on-axis, as seen in Figs. \ref{fig2} (e) and (f). The temporal duration, at full-width half maximum, of the isolated attosecond STOV pulse is $\sim 290$ as.

\begin{figure*}[ht]
\includegraphics[width=0.9\textwidth]{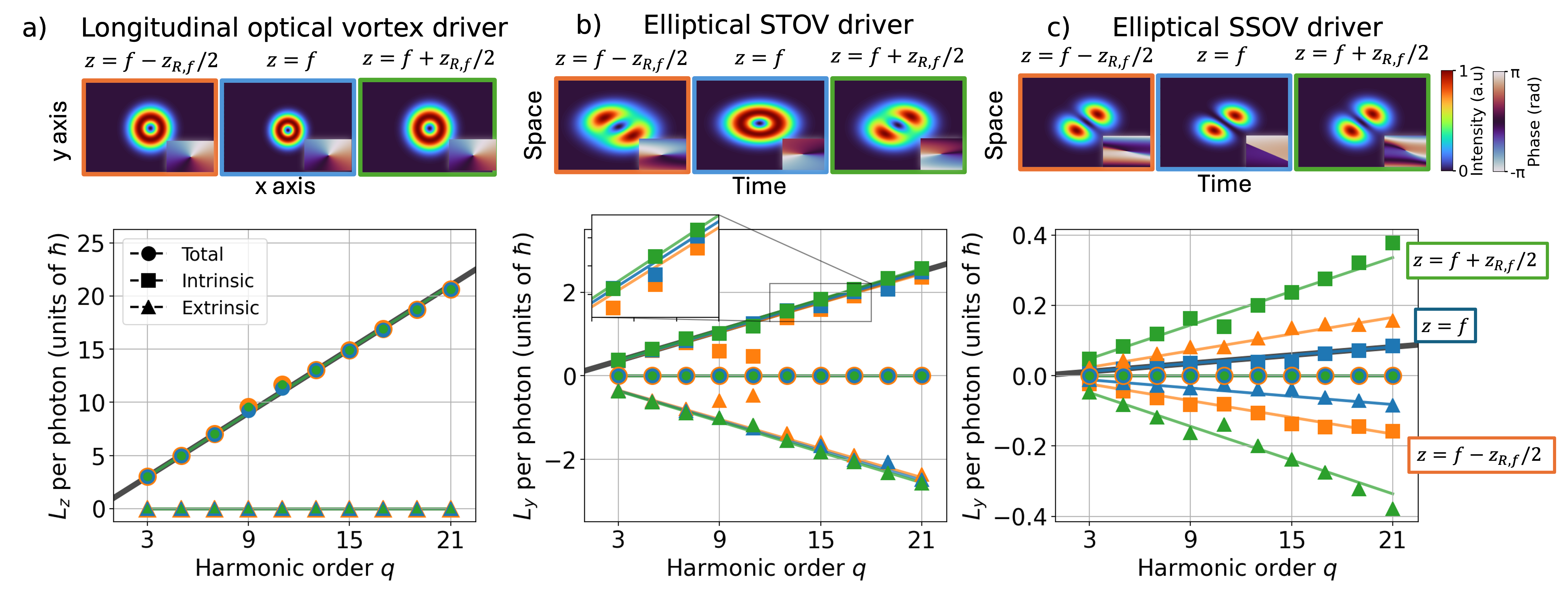}
\caption{Comparative of the harmonic average OAM per photon scaling for (a) LG vortex, (b) STOV, and (c) SSOV as drivers, with the gas-jet placed at distances $f-z_{R,f}/2$ (orange), $f$ (blue) and $f+z_{R,f}/2$ (green), where $z_{R,f}=f^2/z_R$ is the focal Rayleigh range. The intensity and phase of the respective drivers are shown in the top panels. The solid lines are the predictions of the elemental model and the symbols of the advanced numerical simulations. The black line shows the scaling with the intrinsic OAM of the driving field. The parameters of the longitudinal optical vortex  and the SSOV driver (with $\eta = 1$) are the same as in Fig. 1 (b), while the parameters for the STOV driving field corresponds with those in \cite{Martin-Hernandez2025spatiotemporal}.} \label{fig3}
\end{figure*}


Our results question the generalized identification between the topological charge scaling and OAM conservation in photon up-conversion processes. Such association has been also extrapolated to STOVs, even though they present {radically different characteristics to LG vortices, such as} evolving and reversing $\ell$ \cite{porras2023propagation,porras2023transverse,Porras_nano_2025}, and a t-OAM that is not purely intrinsic \cite{Bliokh2012,Bliokh2021,Bliokh2023,porras2023transverse,porras2023propagation}.
Our numerical HHG simulations using LG, STOV and SSOV driving beams show that the average intrinsic OAM per photon ---the only part of the OAM that can be regarded as a property of photons,--- scales with the harmonic order $q$ with generality. However, except in LG-driven HHG, the average intrinsic OAM per photon of the $q^{\rm th}$ harmonic is not $q$ times that of the driver.  


The scaling properties of the average total OAM per photon (round dots) and its intrinsic (squares) and extrinsic (triangles) parts are shown in Fig. 3 (see Supp. Mat., {compiling their mathematical expressions from \cite{porras2023transverse,porras2024clarification}}) are shown in Fig. \ref{fig3}. They are evaluated directly from the harmonic spatiotemporal fields provided by the elemental model (solid lines) and by the advanced numerical simulations (symbols), in HHG driven by (a) LG vortices, (b) STOVs and (c) SSOVs. The gas jet is placed at three positions around the focus ($z=f-z_{R,f}/2$ in orange, $z=f$ in blue, and $z=f+z_{R,f}/2$ in green). All driving fields are shown in the top panels. In panel (a), the total l-OAM is evaluated around the optical axis. In panels (b) and (c) the t-OAM is evaluated around the $y$ axis $(x,z)=0$ (passing through the lens center), and the intrinsic t-OAM is evaluated extracting the extrinsic t-OAM of the STOV/SSOV energy centroid \cite{porras2023transverse,porras2024clarification}. We stress that the intrinsic t-OAM of the driving STOV/SSOV fields at the three gas jet positions is the same as the intrinsic t-OAM is conserved in free propagation with the choice of the energy centroid \cite{porras2023transverse,porras2024clarification,BliokhPhysLettsA}. In the Supp. Mat. we repeat the whole procedure with the choice of the photon centroid \cite{Bliokh2012,Bliokh2021,Bliokh2023}, reaching similar conclusions.

In LG-driven HHG, the harmonic total and intrinsic l-OAM per photon are the same, being $q$ times that of the driver ($\ell=1$), corroborating l-OAM conservation \cite{Hernandez-Garcia2013}.
In STOV/SSOV-driven HHG, the harmonic total t-OAM is null, as that of the drivers, and therefore, it is also conserved [Figs. \ref{fig3}(b) and (c)]. The average intrinsic t-OAM per photon scales with the harmonic order $q$ (and the opposite extrinsic t-OAM) regardless of its topological charge, but it is generally not equal to $q$ times the intrinsic t-OAM per photon of the driver (black straight line), implying its non-conservation. The particular conservation of the intrinsic t-OAM when the gas jet is placed at the focal plane is easily attributable to the symmetries of STOVs/SSOVs at that plane. The general non-conservation is barely observable with the STOV-driver (see the zoomed inset), but it is exacerbated with the SSOV-driver. Thus, the general scaling of the average intrinsic l-OAM and t-OAM per photon with harmonic order $q$ reflects the process of up-conversion of $q$ photons to a harmonic photon in all three cases. However, the same intrinsic and opposite extrinsic t-OAM of the STOV/SSOV drivers are redistributed differently in the highly nonlinear HHG process at the three axial positions, even reversing their signs, depending on the specific spatiotemporal profile at those positions. An axial scan of the gas jet, would result in rather arbitrary, continuous values of the average intrinsic t-OAM per photon, {a situation that differs substantially from LG-driven HHG}.

In conclusion, we demonstrate that attosecond STOV pulses can be created taking into account the non-scaling topological charge behavior of HHG driven by SSOVs. In conjunction with the lighthouse effect, we are able to isolate the attosecond STOV from the generated pulse train. In addition, our results evidence that counting linearly scaling topological charge with harmonic order as a verification of OAM per photon up-conversion, and hence OAM conservation, is only justified for LG drivers, and other cylindrically symmetric vortex beams with l-OAM. With STOVs and SSOVs, $\ell$ and t-OAM per photon up-conversions are generally disconnected phenomena. As a photon up-conversion process, the physical magnitude that generally scales with harmonic order is the intrinsic OAM per photon, but the intrinsic OAM itself is not generally conserved. We stress that there no exist general conservation laws of the intrinsic OAM, only the total OAM (in absence of spin-orbit coupling) of the whole system is conserved in absence of torques \cite{BliokhPhysLettsA}. 
Our proposal opens the route for the generation of EUV wavepackets exhibiting light topologies coupled in space and time at nanometer and attosecond scales. 


\acknowledgements
This project has received funding from the European Research Council (ERC) under the European Union’s Horizon 2020 research and innovation programme (grant agreement No 851201) {and from the Department of Education of the Junta de Castilla y León and FEDER Funds (Escalera de Excelencia CLU-2023-1-02 and grant No. SA108P24)}. C.H.-G. and L.P. acknowledge funding from Ministerio de Ciencia e Innovacion (Grant PID2022-142340NB-I00). M.A.P. acknowledges support from the Spanish Ministry of Science and Innovation, Gobierno de España, under Contract No. PID2021-122711NB-C21.  R.M.-H., L.P. and C.H.-G. Thankfully acknowledge RES resources provided by BSC in MareNostrum 5 {, and CESGA in Finisterrae 3 to 
FI-2024-3-0035.}

\bibliography{main}

\begin{thebibliography}{54}%
\makeatletter
\providecommand \@ifxundefined [1]{%
 \@ifx{#1\undefined}
}%
\providecommand \@ifnum [1]{%
 \ifnum #1\expandafter \@firstoftwo
 \else \expandafter \@secondoftwo
 \fi
}%
\providecommand \@ifx [1]{%
 \ifx #1\expandafter \@firstoftwo
 \else \expandafter \@secondoftwo
 \fi
}%
\providecommand \natexlab [1]{#1}%
\providecommand \enquote  [1]{``#1''}%
\providecommand \bibnamefont  [1]{#1}%
\providecommand \bibfnamefont [1]{#1}%
\providecommand \citenamefont [1]{#1}%
\providecommand \href@noop [0]{\@secondoftwo}%
\providecommand \href [0]{\begingroup \@sanitize@url \@href}%
\providecommand \@href[1]{\@@startlink{#1}\@@href}%
\providecommand \@@href[1]{\endgroup#1\@@endlink}%
\providecommand \@sanitize@url [0]{\catcode `\\12\catcode `\$12\catcode
  `\&12\catcode `\#12\catcode `\^12\catcode `\_12\catcode `\%12\relax}%
\providecommand \@@startlink[1]{}%
\providecommand \@@endlink[0]{}%
\providecommand \url  [0]{\begingroup\@sanitize@url \@url }%
\providecommand \@url [1]{\endgroup\@href {#1}{\urlprefix }}%
\providecommand \urlprefix  [0]{URL }%
\providecommand \Eprint [0]{\href }%
\providecommand \doibase [0]{http://dx.doi.org/}%
\providecommand \selectlanguage [0]{\@gobble}%
\providecommand \bibinfo  [0]{\@secondoftwo}%
\providecommand \bibfield  [0]{\@secondoftwo}%
\providecommand \translation [1]{[#1]}%
\providecommand \BibitemOpen [0]{}%
\providecommand \bibitemStop [0]{}%
\providecommand \bibitemNoStop [0]{.\EOS\space}%
\providecommand \EOS [0]{\spacefactor3000\relax}%
\providecommand \BibitemShut  [1]{\csname bibitem#1\endcsname}%
\let\auto@bib@innerbib\@empty
\bibitem [{\citenamefont {Bliokh}(2012)}]{Bliokh2012}%
  \BibitemOpen
  \bibfield  {author} {\bibinfo {author} {\bibfnamefont {K.}~\bibnamefont
  {Bliokh}},\ }\href@noop {} {\bibfield  {journal} {\bibinfo  {journal}
  {Physical Review A}\ }\textbf {\bibinfo {volume} {86}},\ \bibinfo {pages}
  {033824} (\bibinfo {year} {2012})}\BibitemShut {NoStop}%
\bibitem [{\citenamefont {Bliokh}(2021)}]{Bliokh2021}%
  \BibitemOpen
  \bibfield  {author} {\bibinfo {author} {\bibfnamefont {K.}~\bibnamefont
  {Bliokh}},\ }\href@noop {} {\bibfield  {journal} {\bibinfo  {journal}
  {Physical Review Letters}\ }\textbf {\bibinfo {volume} {126}},\ \bibinfo
  {pages} {243601} (\bibinfo {year} {2021})}\BibitemShut {NoStop}%
\bibitem [{\citenamefont {Bliokh}(2023)}]{Bliokh2023}%
  \BibitemOpen
  \bibfield  {author} {\bibinfo {author} {\bibfnamefont {K.}~\bibnamefont
  {Bliokh}},\ }\href@noop {} {\bibfield  {journal} {\bibinfo  {journal}
  {Physical Review A}\ }\textbf {\bibinfo {volume} {107}},\ \bibinfo {pages}
  {L031501} (\bibinfo {year} {2023})}\BibitemShut {NoStop}%
\bibitem [{\citenamefont {Porras}(2023{\natexlab{a}})}]{porras2023transverse}%
  \BibitemOpen
  \bibfield  {author} {\bibinfo {author} {\bibfnamefont {M.~A.}\ \bibnamefont
  {Porras}},\ }\href@noop {} {\bibfield  {journal} {\bibinfo  {journal} {Prog.
  Electromagn. Res.}\ }\textbf {\bibinfo {volume} {177}},\ \bibinfo {pages}
  {95} (\bibinfo {year} {2023}{\natexlab{a}})}\BibitemShut {NoStop}%
\bibitem [{\citenamefont {Porras}(2024)}]{porras2024clarification}%
  \BibitemOpen
  \bibfield  {author} {\bibinfo {author} {\bibfnamefont {M.}~\bibnamefont
  {Porras}},\ }\href@noop {} {\bibfield  {journal} {\bibinfo  {journal}
  {Journal of Optics}\ }\textbf {\bibinfo {volume} {26}},\ \bibinfo {pages}
  {095601} (\bibinfo {year} {2024})}\BibitemShut {NoStop}%
\bibitem [{\citenamefont {Jhajj}\ \emph {et~al.}(2016)\citenamefont {Jhajj},
  \citenamefont {Larkin}, \citenamefont {Rosenthal}, \citenamefont {Zahedpour},
  \citenamefont {Wahlstrand},\ and\ \citenamefont {Milchberg}}]{Jhajj2016}%
  \BibitemOpen
  \bibfield  {author} {\bibinfo {author} {\bibfnamefont {N.}~\bibnamefont
  {Jhajj}}, \bibinfo {author} {\bibfnamefont {I.}~\bibnamefont {Larkin}},
  \bibinfo {author} {\bibfnamefont {E.~W.}\ \bibnamefont {Rosenthal}}, \bibinfo
  {author} {\bibfnamefont {S.}~\bibnamefont {Zahedpour}}, \bibinfo {author}
  {\bibfnamefont {J.~K.}\ \bibnamefont {Wahlstrand}}, \ and\ \bibinfo {author}
  {\bibfnamefont {H.~M.}\ \bibnamefont {Milchberg}},\ }\href {\doibase
  10.1103/PhysRevX.6.031037} {\bibfield  {journal} {\bibinfo  {journal} {Phys.
  Rev. X}\ }\textbf {\bibinfo {volume} {6}},\ \bibinfo {pages} {031037}
  (\bibinfo {year} {2016})}\BibitemShut {NoStop}%
\bibitem [{\citenamefont {Hancock}\ \emph {et~al.}(2019)\citenamefont
  {Hancock}, \citenamefont {Zahedpour}, \citenamefont {Goffin},\ and\
  \citenamefont {Milchberg}}]{Hancock19}%
  \BibitemOpen
  \bibfield  {author} {\bibinfo {author} {\bibfnamefont {S.~W.}\ \bibnamefont
  {Hancock}}, \bibinfo {author} {\bibfnamefont {S.}~\bibnamefont {Zahedpour}},
  \bibinfo {author} {\bibfnamefont {A.}~\bibnamefont {Goffin}}, \ and\ \bibinfo
  {author} {\bibfnamefont {H.~M.}\ \bibnamefont {Milchberg}},\ }\href {\doibase
  10.1364/OPTICA.6.001547} {\bibfield  {journal} {\bibinfo  {journal} {Optica}\
  }\textbf {\bibinfo {volume} {6}},\ \bibinfo {pages} {1547} (\bibinfo {year}
  {2019})}\BibitemShut {NoStop}%
\bibitem [{\citenamefont {Chong}\ \emph {et~al.}(2020)\citenamefont {Chong},
  \citenamefont {Wan}, \citenamefont {Chen},\ and\ \citenamefont
  {Zhan}}]{chong2020}%
  \BibitemOpen
  \bibfield  {author} {\bibinfo {author} {\bibfnamefont {A.}~\bibnamefont
  {Chong}}, \bibinfo {author} {\bibfnamefont {C.}~\bibnamefont {Wan}}, \bibinfo
  {author} {\bibfnamefont {J.}~\bibnamefont {Chen}}, \ and\ \bibinfo {author}
  {\bibfnamefont {Q.}~\bibnamefont {Zhan}},\ }\href@noop {} {\bibfield
  {journal} {\bibinfo  {journal} {Nature Photonics}\ }\textbf {\bibinfo
  {volume} {14}},\ \bibinfo {pages} {350} (\bibinfo {year} {2020})}\BibitemShut
  {NoStop}%
\bibitem [{\citenamefont {Huang}\ \emph {et~al.}(2024)\citenamefont {Huang},
  \citenamefont {Li}, \citenamefont {Li}, \citenamefont {Zhang}, \citenamefont
  {Lu}, \citenamefont {Dorfman}, \citenamefont {Liu},\ and\ \citenamefont
  {Yao}}]{Huan2024SciAdv}%
  \BibitemOpen
  \bibfield  {author} {\bibinfo {author} {\bibfnamefont {S.}~\bibnamefont
  {Huang}}, \bibinfo {author} {\bibfnamefont {Z.}~\bibnamefont {Li}}, \bibinfo
  {author} {\bibfnamefont {J.}~\bibnamefont {Li}}, \bibinfo {author}
  {\bibfnamefont {N.}~\bibnamefont {Zhang}}, \bibinfo {author} {\bibfnamefont
  {X.}~\bibnamefont {Lu}}, \bibinfo {author} {\bibfnamefont {K.}~\bibnamefont
  {Dorfman}}, \bibinfo {author} {\bibfnamefont {J.}~\bibnamefont {Liu}}, \ and\
  \bibinfo {author} {\bibfnamefont {J.}~\bibnamefont {Yao}},\ }\href {\doibase
  10.1126/sciadv.adn6206} {\bibfield  {journal} {\bibinfo  {journal} {Science
  Advances}\ }\textbf {\bibinfo {volume} {10}},\ \bibinfo {pages} {eadn6206}
  (\bibinfo {year} {2024})},\ \Eprint
  {http://arxiv.org/abs/https://www.science.org/doi/pdf/10.1126/sciadv.adn6206}
  {https://www.science.org/doi/pdf/10.1126/sciadv.adn6206} \BibitemShut
  {NoStop}%
\bibitem [{\citenamefont {Wang}\ \emph {et~al.}(2021)\citenamefont {Wang},
  \citenamefont {Guo}, \citenamefont {Jin}, \citenamefont {Song},\ and\
  \citenamefont {Fan}}]{wang2021}%
  \BibitemOpen
  \bibfield  {author} {\bibinfo {author} {\bibfnamefont {H.}~\bibnamefont
  {Wang}}, \bibinfo {author} {\bibfnamefont {C.}~\bibnamefont {Guo}}, \bibinfo
  {author} {\bibfnamefont {W.}~\bibnamefont {Jin}}, \bibinfo {author}
  {\bibfnamefont {A.~Y.}\ \bibnamefont {Song}}, \ and\ \bibinfo {author}
  {\bibfnamefont {S.}~\bibnamefont {Fan}},\ }\href@noop {} {\bibfield
  {journal} {\bibinfo  {journal} {Optica}\ }\textbf {\bibinfo {volume} {8}},\
  \bibinfo {pages} {966} (\bibinfo {year} {2021})}\BibitemShut {NoStop}%
\bibitem [{\citenamefont {Huang}\ \emph {et~al.}(2023)\citenamefont {Huang},
  \citenamefont {Zhang}, \citenamefont {Lu}, \citenamefont {Liu},\ and\
  \citenamefont {Yao}}]{huang2023spatiotemporal}%
  \BibitemOpen
  \bibfield  {author} {\bibinfo {author} {\bibfnamefont {S.}~\bibnamefont
  {Huang}}, \bibinfo {author} {\bibfnamefont {N.}~\bibnamefont {Zhang}},
  \bibinfo {author} {\bibfnamefont {X.}~\bibnamefont {Lu}}, \bibinfo {author}
  {\bibfnamefont {J.}~\bibnamefont {Liu}}, \ and\ \bibinfo {author}
  {\bibfnamefont {J.}~\bibnamefont {Yao}},\ }\href@noop {} {\bibfield
  {journal} {\bibinfo  {journal} {arXiv preprint arXiv:2305.08407}\ } (\bibinfo
  {year} {2023})}\BibitemShut {NoStop}%
\bibitem [{\citenamefont {Huo}\ \emph {et~al.}(2024)\citenamefont {Huo},
  \citenamefont {Chen}, \citenamefont {Zhang}, \citenamefont {Zhang},
  \citenamefont {Liu}, \citenamefont {Lin}, \citenamefont {Zhang},
  \citenamefont {Chen}, \citenamefont {Lezec}, \citenamefont {Zhu} \emph
  {et~al.}}]{huo2024observation}%
  \BibitemOpen
  \bibfield  {author} {\bibinfo {author} {\bibfnamefont {P.}~\bibnamefont
  {Huo}}, \bibinfo {author} {\bibfnamefont {W.}~\bibnamefont {Chen}}, \bibinfo
  {author} {\bibfnamefont {Z.}~\bibnamefont {Zhang}}, \bibinfo {author}
  {\bibfnamefont {Y.}~\bibnamefont {Zhang}}, \bibinfo {author} {\bibfnamefont
  {M.}~\bibnamefont {Liu}}, \bibinfo {author} {\bibfnamefont {P.}~\bibnamefont
  {Lin}}, \bibinfo {author} {\bibfnamefont {H.}~\bibnamefont {Zhang}}, \bibinfo
  {author} {\bibfnamefont {Z.}~\bibnamefont {Chen}}, \bibinfo {author}
  {\bibfnamefont {H.}~\bibnamefont {Lezec}}, \bibinfo {author} {\bibfnamefont
  {W.}~\bibnamefont {Zhu}},  \emph {et~al.},\ }\href@noop {} {\bibfield
  {journal} {\bibinfo  {journal} {Nature Communications}\ }\textbf {\bibinfo
  {volume} {15}},\ \bibinfo {pages} {3055} (\bibinfo {year}
  {2024})}\BibitemShut {NoStop}%
\bibitem [{\citenamefont {Sun}\ \emph {et~al.}(2024)\citenamefont {Sun},
  \citenamefont {Wang}, \citenamefont {Dong}, \citenamefont {He}, \citenamefont
  {Shi}, \citenamefont {Lv}, \citenamefont {Zhan}, \citenamefont {Leng},
  \citenamefont {Zhuang},\ and\ \citenamefont {Li}}]{Sun2024}%
  \BibitemOpen
  \bibfield  {author} {\bibinfo {author} {\bibfnamefont {F.}~\bibnamefont
  {Sun}}, \bibinfo {author} {\bibfnamefont {W.}~\bibnamefont {Wang}}, \bibinfo
  {author} {\bibfnamefont {H.}~\bibnamefont {Dong}}, \bibinfo {author}
  {\bibfnamefont {J.}~\bibnamefont {He}}, \bibinfo {author} {\bibfnamefont
  {Z.}~\bibnamefont {Shi}}, \bibinfo {author} {\bibfnamefont {Z.}~\bibnamefont
  {Lv}}, \bibinfo {author} {\bibfnamefont {Q.}~\bibnamefont {Zhan}}, \bibinfo
  {author} {\bibfnamefont {Y.}~\bibnamefont {Leng}}, \bibinfo {author}
  {\bibfnamefont {S.}~\bibnamefont {Zhuang}}, \ and\ \bibinfo {author}
  {\bibfnamefont {R.}~\bibnamefont {Li}},\ }\href@noop {} {\bibfield  {journal}
  {\bibinfo  {journal} {Physical Review Research}\ }\textbf {\bibinfo {volume}
  {6}},\ \bibinfo {pages} {013075} (\bibinfo {year} {2024})}\BibitemShut
  {NoStop}%
\bibitem [{\citenamefont {Wan}\ \emph {et~al.}(2022)\citenamefont {Wan},
  \citenamefont {Shen}, \citenamefont {Chong},\ and\ \citenamefont
  {Zhan}}]{wan2022scalar}%
  \BibitemOpen
  \bibfield  {author} {\bibinfo {author} {\bibfnamefont {C.}~\bibnamefont
  {Wan}}, \bibinfo {author} {\bibfnamefont {Y.}~\bibnamefont {Shen}}, \bibinfo
  {author} {\bibfnamefont {A.}~\bibnamefont {Chong}}, \ and\ \bibinfo {author}
  {\bibfnamefont {Q.}~\bibnamefont {Zhan}},\ }\href@noop {} {\bibfield
  {journal} {\bibinfo  {journal} {eLight}\ }\textbf {\bibinfo {volume} {2}},\
  \bibinfo {pages} {22} (\bibinfo {year} {2022})}\BibitemShut {NoStop}%
\bibitem [{\citenamefont {Gui}\ \emph {et~al.}(2021)\citenamefont {Gui},
  \citenamefont {Brooks}, \citenamefont {Kapteyn}, \citenamefont {Murnane},\
  and\ \citenamefont {Liao}}]{gui2021}%
  \BibitemOpen
  \bibfield  {author} {\bibinfo {author} {\bibfnamefont {G.}~\bibnamefont
  {Gui}}, \bibinfo {author} {\bibfnamefont {N.~J.}\ \bibnamefont {Brooks}},
  \bibinfo {author} {\bibfnamefont {H.~C.}\ \bibnamefont {Kapteyn}}, \bibinfo
  {author} {\bibfnamefont {M.~M.}\ \bibnamefont {Murnane}}, \ and\ \bibinfo
  {author} {\bibfnamefont {C.-T.}\ \bibnamefont {Liao}},\ }\href@noop {}
  {\bibfield  {journal} {\bibinfo  {journal} {Nature Photonics}\ }\textbf
  {\bibinfo {volume} {15}},\ \bibinfo {pages} {608} (\bibinfo {year}
  {2021})}\BibitemShut {NoStop}%
\bibitem [{\citenamefont {Hancock}\ \emph {et~al.}(2021)\citenamefont
  {Hancock}, \citenamefont {Zahedpour},\ and\ \citenamefont
  {Milchberg}}]{hancock2021}%
  \BibitemOpen
  \bibfield  {author} {\bibinfo {author} {\bibfnamefont {S.}~\bibnamefont
  {Hancock}}, \bibinfo {author} {\bibfnamefont {S.}~\bibnamefont {Zahedpour}},
  \ and\ \bibinfo {author} {\bibfnamefont {H.}~\bibnamefont {Milchberg}},\
  }\href@noop {} {\bibfield  {journal} {\bibinfo  {journal} {Optica}\ }\textbf
  {\bibinfo {volume} {8}},\ \bibinfo {pages} {594} (\bibinfo {year}
  {2021})}\BibitemShut {NoStop}%
\bibitem [{\citenamefont {Wang}\ \emph {et~al.}(2023)\citenamefont {Wang},
  \citenamefont {Chen}, \citenamefont {Zhang},\ and\ \citenamefont
  {Shen}}]{Wang23third}%
  \BibitemOpen
  \bibfield  {author} {\bibinfo {author} {\bibfnamefont {H.}~\bibnamefont
  {Wang}}, \bibinfo {author} {\bibfnamefont {Y.-Y.}\ \bibnamefont {Chen}},
  \bibinfo {author} {\bibfnamefont {X.}~\bibnamefont {Zhang}}, \ and\ \bibinfo
  {author} {\bibfnamefont {B.}~\bibnamefont {Shen}},\ }\href@noop {} {\bibfield
   {journal} {\bibinfo  {journal} {Opt. Express}\ }\textbf {\bibinfo {volume}
  {31}},\ \bibinfo {pages} {36810} (\bibinfo {year} {2023})}\BibitemShut
  {NoStop}%
\bibitem [{\citenamefont {Gao}\ \emph {et~al.}(2023)\citenamefont {Gao},
  \citenamefont {Zhao}, \citenamefont {Wang}, \citenamefont {Lu}, \citenamefont
  {Zhang}, \citenamefont {Fan},\ and\ \citenamefont {Hu}}]{Gao23}%
  \BibitemOpen
  \bibfield  {author} {\bibinfo {author} {\bibfnamefont {X.}~\bibnamefont
  {Gao}}, \bibinfo {author} {\bibfnamefont {Y.}~\bibnamefont {Zhao}}, \bibinfo
  {author} {\bibfnamefont {J.}~\bibnamefont {Wang}}, \bibinfo {author}
  {\bibfnamefont {Y.}~\bibnamefont {Lu}}, \bibinfo {author} {\bibfnamefont
  {J.}~\bibnamefont {Zhang}}, \bibinfo {author} {\bibfnamefont
  {J.}~\bibnamefont {Fan}}, \ and\ \bibinfo {author} {\bibfnamefont
  {M.}~\bibnamefont {Hu}},\ }\href@noop {} {\bibfield  {journal} {\bibinfo
  {journal} {Chin. Opt. Lett.}\ }\textbf {\bibinfo {volume} {21}},\ \bibinfo
  {pages} {080004} (\bibinfo {year} {2023})}\BibitemShut {NoStop}%
\bibitem [{\citenamefont {McPherson}\ \emph {et~al.}(1987)\citenamefont
  {McPherson}, \citenamefont {Gibson}, \citenamefont {Jara}, \citenamefont
  {Johann}, \citenamefont {Luk}, \citenamefont {McIntyre}, \citenamefont
  {Boyer},\ and\ \citenamefont {Rhodes}}]{McPherson1987}%
  \BibitemOpen
  \bibfield  {author} {\bibinfo {author} {\bibfnamefont {A.}~\bibnamefont
  {McPherson}}, \bibinfo {author} {\bibfnamefont {G.}~\bibnamefont {Gibson}},
  \bibinfo {author} {\bibfnamefont {H.}~\bibnamefont {Jara}}, \bibinfo {author}
  {\bibfnamefont {U.}~\bibnamefont {Johann}}, \bibinfo {author} {\bibfnamefont
  {T.~S.}\ \bibnamefont {Luk}}, \bibinfo {author} {\bibfnamefont {I.~A.}\
  \bibnamefont {McIntyre}}, \bibinfo {author} {\bibfnamefont {K.}~\bibnamefont
  {Boyer}}, \ and\ \bibinfo {author} {\bibfnamefont {C.~K.}\ \bibnamefont
  {Rhodes}},\ }\href {\doibase 10.1364/JOSAB.4.000595} {\bibfield  {journal}
  {\bibinfo  {journal} {J. Opt. Soc. Am. B}\ }\textbf {\bibinfo {volume} {4}},\
  \bibinfo {pages} {595} (\bibinfo {year} {1987})}\BibitemShut {NoStop}%
\bibitem [{\citenamefont {Ferray}\ \emph {et~al.}(1988)\citenamefont {Ferray},
  \citenamefont {L'Huillier}, \citenamefont {Li}, \citenamefont {Lompre},
  \citenamefont {Mainfray},\ and\ \citenamefont {Manus}}]{Ferray1988}%
  \BibitemOpen
  \bibfield  {author} {\bibinfo {author} {\bibfnamefont {M.}~\bibnamefont
  {Ferray}}, \bibinfo {author} {\bibfnamefont {A.}~\bibnamefont {L'Huillier}},
  \bibinfo {author} {\bibfnamefont {X.~F.}\ \bibnamefont {Li}}, \bibinfo
  {author} {\bibfnamefont {L.~A.}\ \bibnamefont {Lompre}}, \bibinfo {author}
  {\bibfnamefont {G.}~\bibnamefont {Mainfray}}, \ and\ \bibinfo {author}
  {\bibfnamefont {C.}~\bibnamefont {Manus}},\ }\href {\doibase
  10.1088/0953-4075/21/3/001} {\bibfield  {journal} {\bibinfo  {journal}
  {Journal of Physics B: Atomic, Molecular and Optical Physics}\ }\textbf
  {\bibinfo {volume} {21}},\ \bibinfo {pages} {L31} (\bibinfo {year}
  {1988})}\BibitemShut {NoStop}%
\bibitem [{\citenamefont {Popmintchev}\ \emph {et~al.}(2012)\citenamefont
  {Popmintchev}, \citenamefont {Chen}, \citenamefont {Popmintchev},
  \citenamefont {Arpin}, \citenamefont {Brown}, \citenamefont {Alisauskas},
  \citenamefont {Andriukaitis}, \citenamefont {Balciunas}, \citenamefont
  {Mücke}, \citenamefont {Pugzlys}, \citenamefont {Baltuska}, \citenamefont
  {Shim}, \citenamefont {Schrauth}, \citenamefont {Gaeta}, \citenamefont
  {Hernández-García}, \citenamefont {Plaja}, \citenamefont {Becker},
  \citenamefont {Jaron-Becker}, \citenamefont {Murnane},\ and\ \citenamefont
  {Kapteyn}}]{Popmintchev2012}%
  \BibitemOpen
  \bibfield  {author} {\bibinfo {author} {\bibfnamefont {T.}~\bibnamefont
  {Popmintchev}}, \bibinfo {author} {\bibfnamefont {M.-C.}\ \bibnamefont
  {Chen}}, \bibinfo {author} {\bibfnamefont {D.}~\bibnamefont {Popmintchev}},
  \bibinfo {author} {\bibfnamefont {P.}~\bibnamefont {Arpin}}, \bibinfo
  {author} {\bibfnamefont {S.}~\bibnamefont {Brown}}, \bibinfo {author}
  {\bibfnamefont {S.}~\bibnamefont {Alisauskas}}, \bibinfo {author}
  {\bibfnamefont {G.}~\bibnamefont {Andriukaitis}}, \bibinfo {author}
  {\bibfnamefont {T.}~\bibnamefont {Balciunas}}, \bibinfo {author}
  {\bibfnamefont {O.~D.}\ \bibnamefont {Mücke}}, \bibinfo {author}
  {\bibfnamefont {A.}~\bibnamefont {Pugzlys}}, \bibinfo {author} {\bibfnamefont
  {A.}~\bibnamefont {Baltuska}}, \bibinfo {author} {\bibfnamefont
  {B.}~\bibnamefont {Shim}}, \bibinfo {author} {\bibfnamefont {S.~E.}\
  \bibnamefont {Schrauth}}, \bibinfo {author} {\bibfnamefont {A.}~\bibnamefont
  {Gaeta}}, \bibinfo {author} {\bibfnamefont {C.}~\bibnamefont
  {Hernández-García}}, \bibinfo {author} {\bibfnamefont {L.}~\bibnamefont
  {Plaja}}, \bibinfo {author} {\bibfnamefont {A.}~\bibnamefont {Becker}},
  \bibinfo {author} {\bibfnamefont {A.}~\bibnamefont {Jaron-Becker}}, \bibinfo
  {author} {\bibfnamefont {M.~M.}\ \bibnamefont {Murnane}}, \ and\ \bibinfo
  {author} {\bibfnamefont {H.~C.}\ \bibnamefont {Kapteyn}},\ }\href {\doibase
  10.1126/science.1218497} {\bibfield  {journal} {\bibinfo  {journal}
  {Science}\ }\textbf {\bibinfo {volume} {336}},\ \bibinfo {pages} {1287}
  (\bibinfo {year} {2012})},\ \Eprint
  {http://arxiv.org/abs/https://www.science.org/doi/pdf/10.1126/science.1218497}
  {https://www.science.org/doi/pdf/10.1126/science.1218497} \BibitemShut
  {NoStop}%
\bibitem [{\citenamefont {Farkas}\ and\ \citenamefont
  {Tóth}(1992)}]{Farkas1992}%
  \BibitemOpen
  \bibfield  {author} {\bibinfo {author} {\bibfnamefont {G.}~\bibnamefont
  {Farkas}}\ and\ \bibinfo {author} {\bibfnamefont {C.}~\bibnamefont {Tóth}},\
  }\href {\doibase https://doi.org/10.1016/0375-9601(92)90534-S} {\bibfield
  {journal} {\bibinfo  {journal} {Physics Letters A}\ }\textbf {\bibinfo
  {volume} {168}},\ \bibinfo {pages} {447} (\bibinfo {year}
  {1992})}\BibitemShut {NoStop}%
\bibitem [{\citenamefont {Antoine}\ \emph {et~al.}(1996)\citenamefont
  {Antoine}, \citenamefont {L'Huillier},\ and\ \citenamefont
  {Lewenstein}}]{Antoine1996}%
  \BibitemOpen
  \bibfield  {author} {\bibinfo {author} {\bibfnamefont {P.}~\bibnamefont
  {Antoine}}, \bibinfo {author} {\bibfnamefont {A.}~\bibnamefont {L'Huillier}},
  \ and\ \bibinfo {author} {\bibfnamefont {M.}~\bibnamefont {Lewenstein}},\
  }\href {\doibase 10.1103/PhysRevLett.77.1234} {\bibfield  {journal} {\bibinfo
   {journal} {Phys. Rev. Lett.}\ }\textbf {\bibinfo {volume} {77}},\ \bibinfo
  {pages} {1234} (\bibinfo {year} {1996})}\BibitemShut {NoStop}%
\bibitem [{\citenamefont {Christov}\ \emph {et~al.}(1997)\citenamefont
  {Christov}, \citenamefont {Murnane},\ and\ \citenamefont
  {Kapteyn}}]{Christov1997}%
  \BibitemOpen
  \bibfield  {author} {\bibinfo {author} {\bibfnamefont {I.~P.}\ \bibnamefont
  {Christov}}, \bibinfo {author} {\bibfnamefont {M.~M.}\ \bibnamefont
  {Murnane}}, \ and\ \bibinfo {author} {\bibfnamefont {H.~C.}\ \bibnamefont
  {Kapteyn}},\ }\href {\doibase 10.1103/PhysRevLett.78.1251} {\bibfield
  {journal} {\bibinfo  {journal} {Phys. Rev. Lett.}\ }\textbf {\bibinfo
  {volume} {78}},\ \bibinfo {pages} {1251} (\bibinfo {year}
  {1997})}\BibitemShut {NoStop}%
\bibitem [{\citenamefont {Paul}\ \emph {et~al.}(2001)\citenamefont {Paul},
  \citenamefont {Toma}, \citenamefont {Breger}, \citenamefont {Mullot},
  \citenamefont {Augé}, \citenamefont {Balcou}, \citenamefont {Muller},\ and\
  \citenamefont {Agostini}}]{Paul2001}%
  \BibitemOpen
  \bibfield  {author} {\bibinfo {author} {\bibfnamefont {P.~M.}\ \bibnamefont
  {Paul}}, \bibinfo {author} {\bibfnamefont {E.~S.}\ \bibnamefont {Toma}},
  \bibinfo {author} {\bibfnamefont {P.}~\bibnamefont {Breger}}, \bibinfo
  {author} {\bibfnamefont {G.}~\bibnamefont {Mullot}}, \bibinfo {author}
  {\bibfnamefont {F.}~\bibnamefont {Augé}}, \bibinfo {author} {\bibfnamefont
  {P.}~\bibnamefont {Balcou}}, \bibinfo {author} {\bibfnamefont {H.~G.}\
  \bibnamefont {Muller}}, \ and\ \bibinfo {author} {\bibfnamefont
  {P.}~\bibnamefont {Agostini}},\ }\href {\doibase 10.1126/science.1059413}
  {\bibfield  {journal} {\bibinfo  {journal} {Science}\ }\textbf {\bibinfo
  {volume} {292}},\ \bibinfo {pages} {1689} (\bibinfo {year} {2001})},\ \Eprint
  {http://arxiv.org/abs/https://www.science.org/doi/pdf/10.1126/science.1059413}
  {https://www.science.org/doi/pdf/10.1126/science.1059413} \BibitemShut
  {NoStop}%
\bibitem [{\citenamefont {Hentschel}\ \emph {et~al.}(2001)\citenamefont
  {Hentschel}, \citenamefont {Kienberger}, \citenamefont {Spielmann},
  \citenamefont {Reider}, \citenamefont {Milosevic}, \citenamefont {Brabec},
  \citenamefont {Corkum}, \citenamefont {Heinzmann}, \citenamefont {Drescher},\
  and\ \citenamefont {Krausz}}]{Hentschel2001}%
  \BibitemOpen
  \bibfield  {author} {\bibinfo {author} {\bibfnamefont {M.}~\bibnamefont
  {Hentschel}}, \bibinfo {author} {\bibfnamefont {R.}~\bibnamefont
  {Kienberger}}, \bibinfo {author} {\bibfnamefont {C.}~\bibnamefont
  {Spielmann}}, \bibinfo {author} {\bibfnamefont {G.~A.}\ \bibnamefont
  {Reider}}, \bibinfo {author} {\bibfnamefont {N.}~\bibnamefont {Milosevic}},
  \bibinfo {author} {\bibfnamefont {T.}~\bibnamefont {Brabec}}, \bibinfo
  {author} {\bibfnamefont {P.}~\bibnamefont {Corkum}}, \bibinfo {author}
  {\bibfnamefont {U.}~\bibnamefont {Heinzmann}}, \bibinfo {author}
  {\bibfnamefont {M.}~\bibnamefont {Drescher}}, \ and\ \bibinfo {author}
  {\bibfnamefont {F.}~\bibnamefont {Krausz}},\ }\href {\doibase
  10.1038/35107000} {\bibfield  {journal} {\bibinfo  {journal} {Nature}\
  }\textbf {\bibinfo {volume} {414}},\ \bibinfo {pages} {509} (\bibinfo {year}
  {2001})}\BibitemShut {NoStop}%
\bibitem [{\citenamefont {Hern{\'a}ndez-Garc{\'\i}a}\ \emph
  {et~al.}(2013)\citenamefont {Hern{\'a}ndez-Garc{\'\i}a}, \citenamefont
  {Pic{\'o}n}, \citenamefont {San~Rom{\'a}n},\ and\ \citenamefont
  {Plaja}}]{Hernandez-Garcia2013}%
  \BibitemOpen
  \bibfield  {author} {\bibinfo {author} {\bibfnamefont {C.}~\bibnamefont
  {Hern{\'a}ndez-Garc{\'\i}a}}, \bibinfo {author} {\bibfnamefont
  {A.}~\bibnamefont {Pic{\'o}n}}, \bibinfo {author} {\bibfnamefont
  {J.}~\bibnamefont {San~Rom{\'a}n}}, \ and\ \bibinfo {author} {\bibfnamefont
  {L.}~\bibnamefont {Plaja}},\ }\href@noop {} {\bibfield  {journal} {\bibinfo
  {journal} {Physical review letters}\ }\textbf {\bibinfo {volume} {111}},\
  \bibinfo {pages} {083602} (\bibinfo {year} {2013})}\BibitemShut {NoStop}%
\bibitem [{\citenamefont {Gariepy}\ \emph {et~al.}(2014)\citenamefont
  {Gariepy}, \citenamefont {Leach}, \citenamefont {Kim}, \citenamefont
  {Hammond}, \citenamefont {Frumker}, \citenamefont {Boyd},\ and\ \citenamefont
  {Corkum}}]{Gariepy2014}%
  \BibitemOpen
  \bibfield  {author} {\bibinfo {author} {\bibfnamefont {G.}~\bibnamefont
  {Gariepy}}, \bibinfo {author} {\bibfnamefont {J.}~\bibnamefont {Leach}},
  \bibinfo {author} {\bibfnamefont {K.~T.}\ \bibnamefont {Kim}}, \bibinfo
  {author} {\bibfnamefont {T.~J.}\ \bibnamefont {Hammond}}, \bibinfo {author}
  {\bibfnamefont {E.}~\bibnamefont {Frumker}}, \bibinfo {author} {\bibfnamefont
  {R.~W.}\ \bibnamefont {Boyd}}, \ and\ \bibinfo {author} {\bibfnamefont
  {P.~B.}\ \bibnamefont {Corkum}},\ }\href@noop {} {\bibfield  {journal}
  {\bibinfo  {journal} {Physical review letters}\ }\textbf {\bibinfo {volume}
  {113}},\ \bibinfo {pages} {153901} (\bibinfo {year} {2014})}\BibitemShut
  {NoStop}%
\bibitem [{\citenamefont {G{\'e}neaux}\ \emph {et~al.}(2016)\citenamefont
  {G{\'e}neaux}, \citenamefont {Camper}, \citenamefont {Auguste}, \citenamefont
  {Gobert}, \citenamefont {Caillat}, \citenamefont {Ta{\"\i}eb},\ and\
  \citenamefont {Ruchon}}]{Geneaux2016}%
  \BibitemOpen
  \bibfield  {author} {\bibinfo {author} {\bibfnamefont {R.}~\bibnamefont
  {G{\'e}neaux}}, \bibinfo {author} {\bibfnamefont {A.}~\bibnamefont {Camper}},
  \bibinfo {author} {\bibfnamefont {T.}~\bibnamefont {Auguste}}, \bibinfo
  {author} {\bibfnamefont {O.}~\bibnamefont {Gobert}}, \bibinfo {author}
  {\bibfnamefont {J.}~\bibnamefont {Caillat}}, \bibinfo {author} {\bibfnamefont
  {R.}~\bibnamefont {Ta{\"\i}eb}}, \ and\ \bibinfo {author} {\bibfnamefont
  {T.}~\bibnamefont {Ruchon}},\ }\href@noop {} {\bibfield  {journal} {\bibinfo
  {journal} {Nature Communications}\ }\textbf {\bibinfo {volume} {7}},\
  \bibinfo {pages} {12583} (\bibinfo {year} {2016})}\BibitemShut {NoStop}%
\bibitem [{\citenamefont {Pandey}\ \emph {et~al.}(2022)\citenamefont {Pandey},
  \citenamefont {de~las Heras}, \citenamefont {Larrieu}, \citenamefont
  {San~Rom{\'a}n}, \citenamefont {Serrano}, \citenamefont {Plaja},
  \citenamefont {Baynard}, \citenamefont {Pittman}, \citenamefont {Dovillaire},
  \citenamefont {Kazamias} \emph {et~al.}}]{Pandey2022}%
  \BibitemOpen
  \bibfield  {author} {\bibinfo {author} {\bibfnamefont {A.~K.}\ \bibnamefont
  {Pandey}}, \bibinfo {author} {\bibfnamefont {A.}~\bibnamefont {de~las
  Heras}}, \bibinfo {author} {\bibfnamefont {T.}~\bibnamefont {Larrieu}},
  \bibinfo {author} {\bibfnamefont {J.}~\bibnamefont {San~Rom{\'a}n}}, \bibinfo
  {author} {\bibfnamefont {J.}~\bibnamefont {Serrano}}, \bibinfo {author}
  {\bibfnamefont {L.}~\bibnamefont {Plaja}}, \bibinfo {author} {\bibfnamefont
  {E.}~\bibnamefont {Baynard}}, \bibinfo {author} {\bibfnamefont
  {M.}~\bibnamefont {Pittman}}, \bibinfo {author} {\bibfnamefont
  {G.}~\bibnamefont {Dovillaire}}, \bibinfo {author} {\bibfnamefont
  {S.}~\bibnamefont {Kazamias}},  \emph {et~al.},\ }\href@noop {} {\bibfield
  {journal} {\bibinfo  {journal} {ACS Photonics}\ }\textbf {\bibinfo {volume}
  {9}},\ \bibinfo {pages} {944} (\bibinfo {year} {2022})}\BibitemShut {NoStop}%
\bibitem [{\citenamefont {Turpin}\ \emph {et~al.}(2017)\citenamefont {Turpin},
  \citenamefont {Rego}, \citenamefont {Picón}, \citenamefont {Román},\ and\
  \citenamefont {Hernández-García}}]{Turpin2017}%
  \BibitemOpen
  \bibfield  {author} {\bibinfo {author} {\bibfnamefont {A.}~\bibnamefont
  {Turpin}}, \bibinfo {author} {\bibfnamefont {L.}~\bibnamefont {Rego}},
  \bibinfo {author} {\bibfnamefont {A.}~\bibnamefont {Picón}}, \bibinfo
  {author} {\bibfnamefont {J.~S.}\ \bibnamefont {Román}}, \ and\ \bibinfo
  {author} {\bibfnamefont {C.}~\bibnamefont {Hernández-García}},\ }\href
  {\doibase 10.1038/srep43888} {\bibfield  {journal} {\bibinfo  {journal}
  {Scientific Reports}\ }\textbf {\bibinfo {volume} {7}},\ \bibinfo {pages}
  {43888} (\bibinfo {year} {2017})}\BibitemShut {NoStop}%
\bibitem [{\citenamefont {Gauthier}\ \emph {et~al.}(2017)\citenamefont
  {Gauthier}, \citenamefont {Ribi{\v{c}}}, \citenamefont {Adhikary},
  \citenamefont {Camper}, \citenamefont {Chappuis}, \citenamefont {Cucini},
  \citenamefont {DiMauro}, \citenamefont {Dovillaire}, \citenamefont
  {Frassetto}, \citenamefont {G{\'e}neaux} \emph {et~al.}}]{Gauthier2017}%
  \BibitemOpen
  \bibfield  {author} {\bibinfo {author} {\bibfnamefont {D.}~\bibnamefont
  {Gauthier}}, \bibinfo {author} {\bibfnamefont {P.~R.}\ \bibnamefont
  {Ribi{\v{c}}}}, \bibinfo {author} {\bibfnamefont {G.}~\bibnamefont
  {Adhikary}}, \bibinfo {author} {\bibfnamefont {A.}~\bibnamefont {Camper}},
  \bibinfo {author} {\bibfnamefont {C.}~\bibnamefont {Chappuis}}, \bibinfo
  {author} {\bibfnamefont {R.}~\bibnamefont {Cucini}}, \bibinfo {author}
  {\bibfnamefont {L.}~\bibnamefont {DiMauro}}, \bibinfo {author} {\bibfnamefont
  {G.}~\bibnamefont {Dovillaire}}, \bibinfo {author} {\bibfnamefont
  {F.}~\bibnamefont {Frassetto}}, \bibinfo {author} {\bibfnamefont
  {R.}~\bibnamefont {G{\'e}neaux}},  \emph {et~al.},\ }\href@noop {} {\bibfield
   {journal} {\bibinfo  {journal} {Nature Communications}\ }\textbf {\bibinfo
  {volume} {8}},\ \bibinfo {pages} {14971} (\bibinfo {year}
  {2017})}\BibitemShut {NoStop}%
\bibitem [{\citenamefont {Kong}\ \emph {et~al.}(2017)\citenamefont {Kong},
  \citenamefont {Zhang}, \citenamefont {Bouchard}, \citenamefont {Li},
  \citenamefont {Brown}, \citenamefont {Ko}, \citenamefont {Hammond},
  \citenamefont {Arissian}, \citenamefont {Boyd}, \citenamefont {Karimi} \emph
  {et~al.}}]{Kong2017}%
  \BibitemOpen
  \bibfield  {author} {\bibinfo {author} {\bibfnamefont {F.}~\bibnamefont
  {Kong}}, \bibinfo {author} {\bibfnamefont {C.}~\bibnamefont {Zhang}},
  \bibinfo {author} {\bibfnamefont {F.}~\bibnamefont {Bouchard}}, \bibinfo
  {author} {\bibfnamefont {Z.}~\bibnamefont {Li}}, \bibinfo {author}
  {\bibfnamefont {G.~G.}\ \bibnamefont {Brown}}, \bibinfo {author}
  {\bibfnamefont {D.~H.}\ \bibnamefont {Ko}}, \bibinfo {author} {\bibfnamefont
  {T.}~\bibnamefont {Hammond}}, \bibinfo {author} {\bibfnamefont
  {L.}~\bibnamefont {Arissian}}, \bibinfo {author} {\bibfnamefont {R.~W.}\
  \bibnamefont {Boyd}}, \bibinfo {author} {\bibfnamefont {E.}~\bibnamefont
  {Karimi}},  \emph {et~al.},\ }\href@noop {} {\bibfield  {journal} {\bibinfo
  {journal} {Nature communications}\ }\textbf {\bibinfo {volume} {8}},\
  \bibinfo {pages} {14970} (\bibinfo {year} {2017})}\BibitemShut {NoStop}%
\bibitem [{\citenamefont {Dorney}\ \emph {et~al.}(2019)\citenamefont {Dorney},
  \citenamefont {Rego}, \citenamefont {Brooks}, \citenamefont {San~Rom{\'a}n},
  \citenamefont {Liao}, \citenamefont {Ellis}, \citenamefont {Zusin},
  \citenamefont {Gentry}, \citenamefont {Nguyen}, \citenamefont {Shaw},
  \citenamefont {Pic{\'o}n}, \citenamefont {Plaja}, \citenamefont {Kapteyn},
  \citenamefont {Murnane},\ and\ \citenamefont
  {Hern{\'a}ndez-Garc{\'\i}a}}]{dorney_2019_SAMOAMHHG}%
  \BibitemOpen
  \bibfield  {author} {\bibinfo {author} {\bibfnamefont {K.~M.}\ \bibnamefont
  {Dorney}}, \bibinfo {author} {\bibfnamefont {L.}~\bibnamefont {Rego}},
  \bibinfo {author} {\bibfnamefont {N.~J.}\ \bibnamefont {Brooks}}, \bibinfo
  {author} {\bibfnamefont {J.}~\bibnamefont {San~Rom{\'a}n}}, \bibinfo {author}
  {\bibfnamefont {C.-T.}\ \bibnamefont {Liao}}, \bibinfo {author}
  {\bibfnamefont {J.~L.}\ \bibnamefont {Ellis}}, \bibinfo {author}
  {\bibfnamefont {D.}~\bibnamefont {Zusin}}, \bibinfo {author} {\bibfnamefont
  {C.}~\bibnamefont {Gentry}}, \bibinfo {author} {\bibfnamefont {Q.~L.}\
  \bibnamefont {Nguyen}}, \bibinfo {author} {\bibfnamefont {J.~M.}\
  \bibnamefont {Shaw}}, \bibinfo {author} {\bibfnamefont {A.}~\bibnamefont
  {Pic{\'o}n}}, \bibinfo {author} {\bibfnamefont {L.}~\bibnamefont {Plaja}},
  \bibinfo {author} {\bibfnamefont {H.~C.}\ \bibnamefont {Kapteyn}}, \bibinfo
  {author} {\bibfnamefont {M.~M.}\ \bibnamefont {Murnane}}, \ and\ \bibinfo
  {author} {\bibfnamefont {C.}~\bibnamefont {Hern{\'a}ndez-Garc{\'\i}a}},\
  }\href {\doibase 10.1038/s41566-018-0304-3} {\bibfield  {journal} {\bibinfo
  {journal} {Nature Photonics}\ }\textbf {\bibinfo {volume} {13}},\ \bibinfo
  {pages} {123} (\bibinfo {year} {2019})}\BibitemShut {NoStop}%
\bibitem [{\citenamefont {Pisanty}\ \emph {et~al.}(2019)\citenamefont
  {Pisanty}, \citenamefont {Rego}, \citenamefont {San~Rom\'an}, \citenamefont
  {Pic\'on}, \citenamefont {Dorney}, \citenamefont {Kapteyn}, \citenamefont
  {Murnane}, \citenamefont {Plaja}, \citenamefont {Lewenstein},\ and\
  \citenamefont {Hern\'andez-Garc\'{\i}a}}]{Pisanty2019}%
  \BibitemOpen
  \bibfield  {author} {\bibinfo {author} {\bibfnamefont {E.}~\bibnamefont
  {Pisanty}}, \bibinfo {author} {\bibfnamefont {L.}~\bibnamefont {Rego}},
  \bibinfo {author} {\bibfnamefont {J.}~\bibnamefont {San~Rom\'an}}, \bibinfo
  {author} {\bibfnamefont {A.}~\bibnamefont {Pic\'on}}, \bibinfo {author}
  {\bibfnamefont {K.~M.}\ \bibnamefont {Dorney}}, \bibinfo {author}
  {\bibfnamefont {H.~C.}\ \bibnamefont {Kapteyn}}, \bibinfo {author}
  {\bibfnamefont {M.~M.}\ \bibnamefont {Murnane}}, \bibinfo {author}
  {\bibfnamefont {L.}~\bibnamefont {Plaja}}, \bibinfo {author} {\bibfnamefont
  {M.}~\bibnamefont {Lewenstein}}, \ and\ \bibinfo {author} {\bibfnamefont
  {C.}~\bibnamefont {Hern\'andez-Garc\'{\i}a}},\ }\href {\doibase
  10.1103/PhysRevLett.122.203201} {\bibfield  {journal} {\bibinfo  {journal}
  {Phys. Rev. Lett.}\ }\textbf {\bibinfo {volume} {122}},\ \bibinfo {pages}
  {203201} (\bibinfo {year} {2019})}\BibitemShut {NoStop}%
\bibitem [{\citenamefont {de~las Heras}\ \emph {et~al.}(2022)\citenamefont
  {de~las Heras}, \citenamefont {Pandey}, \citenamefont {Rom\'{a}n},
  \citenamefont {Serrano}, \citenamefont {Baynard}, \citenamefont {Dovillaire},
  \citenamefont {Pittman}, \citenamefont {Durfee}, \citenamefont {Plaja},
  \citenamefont {Kazamias}, \citenamefont {Guilbaud},\ and\ \citenamefont
  {Hern\'{a}ndez-Garc\'{i}a}}]{delasHeras_2022}%
  \BibitemOpen
  \bibfield  {author} {\bibinfo {author} {\bibfnamefont {A.}~\bibnamefont
  {de~las Heras}}, \bibinfo {author} {\bibfnamefont {A.~K.}\ \bibnamefont
  {Pandey}}, \bibinfo {author} {\bibfnamefont {J.~S.}\ \bibnamefont
  {Rom\'{a}n}}, \bibinfo {author} {\bibfnamefont {J.}~\bibnamefont {Serrano}},
  \bibinfo {author} {\bibfnamefont {E.}~\bibnamefont {Baynard}}, \bibinfo
  {author} {\bibfnamefont {G.}~\bibnamefont {Dovillaire}}, \bibinfo {author}
  {\bibfnamefont {M.}~\bibnamefont {Pittman}}, \bibinfo {author} {\bibfnamefont
  {C.~G.}\ \bibnamefont {Durfee}}, \bibinfo {author} {\bibfnamefont
  {L.}~\bibnamefont {Plaja}}, \bibinfo {author} {\bibfnamefont
  {S.}~\bibnamefont {Kazamias}}, \bibinfo {author} {\bibfnamefont
  {O.}~\bibnamefont {Guilbaud}}, \ and\ \bibinfo {author} {\bibfnamefont
  {C.}~\bibnamefont {Hern\'{a}ndez-Garc\'{i}a}},\ }\href {\doibase
  10.1364/OPTICA.442304} {\bibfield  {journal} {\bibinfo  {journal} {Optica}\
  }\textbf {\bibinfo {volume} {9}},\ \bibinfo {pages} {71} (\bibinfo {year}
  {2022})}\BibitemShut {NoStop}%
\bibitem [{\citenamefont {Luttmann}\ \emph {et~al.}(2023)\citenamefont
  {Luttmann}, \citenamefont {Vimal}, \citenamefont {Guer}, \citenamefont
  {Hergott}, \citenamefont {Khoury}, \citenamefont {Hern{\'a}ndez-Garc{\'\i}a},
  \citenamefont {Pisanty},\ and\ \citenamefont
  {Ruchon}}]{luttmann2023nonlinear}%
  \BibitemOpen
  \bibfield  {author} {\bibinfo {author} {\bibfnamefont {M.}~\bibnamefont
  {Luttmann}}, \bibinfo {author} {\bibfnamefont {M.}~\bibnamefont {Vimal}},
  \bibinfo {author} {\bibfnamefont {M.}~\bibnamefont {Guer}}, \bibinfo {author}
  {\bibfnamefont {J.-F.}\ \bibnamefont {Hergott}}, \bibinfo {author}
  {\bibfnamefont {A.~Z.}\ \bibnamefont {Khoury}}, \bibinfo {author}
  {\bibfnamefont {C.}~\bibnamefont {Hern{\'a}ndez-Garc{\'\i}a}}, \bibinfo
  {author} {\bibfnamefont {E.}~\bibnamefont {Pisanty}}, \ and\ \bibinfo
  {author} {\bibfnamefont {T.}~\bibnamefont {Ruchon}},\ }\href@noop {}
  {\bibfield  {journal} {\bibinfo  {journal} {Science Advances}\ }\textbf
  {\bibinfo {volume} {9}},\ \bibinfo {pages} {eadf3486} (\bibinfo {year}
  {2023})}\BibitemShut {NoStop}%
\bibitem [{\citenamefont {de~las Heras}\ \emph {et~al.}(2024)\citenamefont
  {de~las Heras}, \citenamefont {Schmidt}, \citenamefont {San~Rom{\'a}n},
  \citenamefont {Serrano}, \citenamefont {Barolak}, \citenamefont {Ivanic},
  \citenamefont {Clarke}, \citenamefont {Westlake}, \citenamefont {Adams},
  \citenamefont {Plaja} \emph {et~al.}}]{de2024attosecond}%
  \BibitemOpen
  \bibfield  {author} {\bibinfo {author} {\bibfnamefont {A.}~\bibnamefont
  {de~las Heras}}, \bibinfo {author} {\bibfnamefont {D.}~\bibnamefont
  {Schmidt}}, \bibinfo {author} {\bibfnamefont {J.}~\bibnamefont
  {San~Rom{\'a}n}}, \bibinfo {author} {\bibfnamefont {J.}~\bibnamefont
  {Serrano}}, \bibinfo {author} {\bibfnamefont {J.}~\bibnamefont {Barolak}},
  \bibinfo {author} {\bibfnamefont {B.}~\bibnamefont {Ivanic}}, \bibinfo
  {author} {\bibfnamefont {C.}~\bibnamefont {Clarke}}, \bibinfo {author}
  {\bibfnamefont {N.}~\bibnamefont {Westlake}}, \bibinfo {author}
  {\bibfnamefont {D.~E.}\ \bibnamefont {Adams}}, \bibinfo {author}
  {\bibfnamefont {L.}~\bibnamefont {Plaja}},  \emph {et~al.},\ }\href@noop {}
  {\bibfield  {journal} {\bibinfo  {journal} {Optica}\ }\textbf {\bibinfo
  {volume} {11}},\ \bibinfo {pages} {1085} (\bibinfo {year}
  {2024})}\BibitemShut {NoStop}%
\bibitem [{\citenamefont {Fang}\ \emph {et~al.}(2021)\citenamefont {Fang},
  \citenamefont {Lu},\ and\ \citenamefont {Liu}}]{fang2021}%
  \BibitemOpen
  \bibfield  {author} {\bibinfo {author} {\bibfnamefont {Y.}~\bibnamefont
  {Fang}}, \bibinfo {author} {\bibfnamefont {S.}~\bibnamefont {Lu}}, \ and\
  \bibinfo {author} {\bibfnamefont {Y.}~\bibnamefont {Liu}},\ }\href@noop {}
  {\bibfield  {journal} {\bibinfo  {journal} {Physical Review Letters}\
  }\textbf {\bibinfo {volume} {127}},\ \bibinfo {pages} {273901} (\bibinfo
  {year} {2021})}\BibitemShut {NoStop}%
\bibitem [{\citenamefont {Martín-Hernández}\ \emph
  {et~al.}(2025)\citenamefont {Martín-Hernández}, \citenamefont {Gui},
  \citenamefont {Plaja}, \citenamefont {Kapteyn}, \citenamefont {Murnane},
  \citenamefont {Liao}, \citenamefont {Porras},\ and\ \citenamefont
  {Hernández-García}}]{Martin-Hernandez2025spatiotemporal}%
  \BibitemOpen
  \bibfield  {author} {\bibinfo {author} {\bibfnamefont {R.}~\bibnamefont
  {Martín-Hernández}}, \bibinfo {author} {\bibfnamefont {G.}~\bibnamefont
  {Gui}}, \bibinfo {author} {\bibfnamefont {L.}~\bibnamefont {Plaja}}, \bibinfo
  {author} {\bibfnamefont {H.~C.}\ \bibnamefont {Kapteyn}}, \bibinfo {author}
  {\bibfnamefont {M.~M.}\ \bibnamefont {Murnane}}, \bibinfo {author}
  {\bibfnamefont {C.-T.}\ \bibnamefont {Liao}}, \bibinfo {author}
  {\bibfnamefont {M.~A.}\ \bibnamefont {Porras}}, \ and\ \bibinfo {author}
  {\bibfnamefont {C.}~\bibnamefont {Hernández-García}},\ }\href@noop {}
  {\bibfield  {journal} {\bibinfo  {journal} {Nature Photonics, in press. arXiv
  preprint arXiv:2412.01716}\ } (\bibinfo {year} {2025})}\BibitemShut {NoStop}%
\bibitem [{\citenamefont {Vincenti}\ and\ \citenamefont
  {Qu{\'e}r{\'e}}(2012)}]{vincenti2012attosecond}%
  \BibitemOpen
  \bibfield  {author} {\bibinfo {author} {\bibfnamefont {H.}~\bibnamefont
  {Vincenti}}\ and\ \bibinfo {author} {\bibfnamefont {F.}~\bibnamefont
  {Qu{\'e}r{\'e}}},\ }\href@noop {} {\bibfield  {journal} {\bibinfo  {journal}
  {Physical review letters}\ }\textbf {\bibinfo {volume} {108}},\ \bibinfo
  {pages} {113904} (\bibinfo {year} {2012})}\BibitemShut {NoStop}%
\bibitem [{\citenamefont {Wheeler}\ \emph {et~al.}(2012)\citenamefont
  {Wheeler}, \citenamefont {Borot}, \citenamefont {Monchoc{\'e}}, \citenamefont
  {Vincenti}, \citenamefont {Ricci}, \citenamefont {Malvache}, \citenamefont
  {Lopez-Martens},\ and\ \citenamefont
  {Qu{\'e}r{\'e}}}]{wheeler2012attosecond}%
  \BibitemOpen
  \bibfield  {author} {\bibinfo {author} {\bibfnamefont {J.~A.}\ \bibnamefont
  {Wheeler}}, \bibinfo {author} {\bibfnamefont {A.}~\bibnamefont {Borot}},
  \bibinfo {author} {\bibfnamefont {S.}~\bibnamefont {Monchoc{\'e}}}, \bibinfo
  {author} {\bibfnamefont {H.}~\bibnamefont {Vincenti}}, \bibinfo {author}
  {\bibfnamefont {A.}~\bibnamefont {Ricci}}, \bibinfo {author} {\bibfnamefont
  {A.}~\bibnamefont {Malvache}}, \bibinfo {author} {\bibfnamefont
  {R.}~\bibnamefont {Lopez-Martens}}, \ and\ \bibinfo {author} {\bibfnamefont
  {F.}~\bibnamefont {Qu{\'e}r{\'e}}},\ }\href@noop {} {\bibfield  {journal}
  {\bibinfo  {journal} {Nature Photonics}\ }\textbf {\bibinfo {volume} {6}},\
  \bibinfo {pages} {829} (\bibinfo {year} {2012})}\BibitemShut {NoStop}%
\bibitem [{\citenamefont {Kim}\ \emph {et~al.}(2013)\citenamefont {Kim},
  \citenamefont {Zhang}, \citenamefont {Ruchon}, \citenamefont {Hergott},
  \citenamefont {Auguste}, \citenamefont {Villeneuve}, \citenamefont {Corkum},\
  and\ \citenamefont {Qu{\'e}r{\'e}}}]{kim2013photonic}%
  \BibitemOpen
  \bibfield  {author} {\bibinfo {author} {\bibfnamefont {K.}~\bibnamefont
  {Kim}}, \bibinfo {author} {\bibfnamefont {C.}~\bibnamefont {Zhang}}, \bibinfo
  {author} {\bibfnamefont {T.}~\bibnamefont {Ruchon}}, \bibinfo {author}
  {\bibfnamefont {J.}~\bibnamefont {Hergott}}, \bibinfo {author} {\bibfnamefont
  {T.}~\bibnamefont {Auguste}}, \bibinfo {author} {\bibfnamefont
  {D.}~\bibnamefont {Villeneuve}}, \bibinfo {author} {\bibfnamefont
  {P.}~\bibnamefont {Corkum}}, \ and\ \bibinfo {author} {\bibfnamefont
  {F.}~\bibnamefont {Qu{\'e}r{\'e}}},\ }\href@noop {} {\bibfield  {journal}
  {\bibinfo  {journal} {Nature Photonics}\ }\textbf {\bibinfo {volume} {7}},\
  \bibinfo {pages} {651} (\bibinfo {year} {2013})}\BibitemShut {NoStop}%
\bibitem [{\citenamefont {Hammond}\ \emph {et~al.}(2016)\citenamefont
  {Hammond}, \citenamefont {Brown}, \citenamefont {Kim}, \citenamefont
  {Villeneuve},\ and\ \citenamefont {Corkum}}]{hammond2016attosecond}%
  \BibitemOpen
  \bibfield  {author} {\bibinfo {author} {\bibfnamefont {T.}~\bibnamefont
  {Hammond}}, \bibinfo {author} {\bibfnamefont {G.~G.}\ \bibnamefont {Brown}},
  \bibinfo {author} {\bibfnamefont {K.~T.}\ \bibnamefont {Kim}}, \bibinfo
  {author} {\bibfnamefont {D.}~\bibnamefont {Villeneuve}}, \ and\ \bibinfo
  {author} {\bibfnamefont {P.}~\bibnamefont {Corkum}},\ }\href@noop {}
  {\bibfield  {journal} {\bibinfo  {journal} {Nature Photonics}\ }\textbf
  {\bibinfo {volume} {10}},\ \bibinfo {pages} {171} (\bibinfo {year}
  {2016})}\BibitemShut {NoStop}%
\bibitem [{\citenamefont {O'Neil}\ \emph {et~al.}(2002)\citenamefont {O'Neil},
  \citenamefont {MacVicar}, \citenamefont {Allen},\ and\ \citenamefont
  {Padgett}}]{ONeil}%
  \BibitemOpen
  \bibfield  {author} {\bibinfo {author} {\bibfnamefont {A.}~\bibnamefont
  {O'Neil}}, \bibinfo {author} {\bibfnamefont {I.}~\bibnamefont {MacVicar}},
  \bibinfo {author} {\bibfnamefont {L.}~\bibnamefont {Allen}}, \ and\ \bibinfo
  {author} {\bibfnamefont {M.}~\bibnamefont {Padgett}},\ }\href@noop {}
  {\bibfield  {journal} {\bibinfo  {journal} {Physical Review Letters}\
  }\textbf {\bibinfo {volume} {88}},\ \bibinfo {pages} {053601} (\bibinfo
  {year} {2002})}\BibitemShut {NoStop}%
\bibitem [{\citenamefont {Porras}(2023{\natexlab{b}})}]{porras2023propagation}%
  \BibitemOpen
  \bibfield  {author} {\bibinfo {author} {\bibfnamefont {M.~A.}\ \bibnamefont
  {Porras}},\ }\href@noop {} {\bibfield  {journal} {\bibinfo  {journal} {Optics
  Letters}\ }\textbf {\bibinfo {volume} {48}},\ \bibinfo {pages} {367}
  (\bibinfo {year} {2023}{\natexlab{b}})}\BibitemShut {NoStop}%
\bibitem [{\citenamefont {Porras}(2025)}]{Porras_nano_2025}%
  \BibitemOpen
  \bibfield  {author} {\bibinfo {author} {\bibfnamefont {M.~A.}\ \bibnamefont
  {Porras}},\ }\href {\doibase 10.1515/nanoph-2024-0544} {\bibfield  {journal}
  {\bibinfo  {journal} {Nanophotonics}\ } (\bibinfo {year} {2025}),\
  10.1515/nanoph-2024-0544}\BibitemShut {NoStop}%
\bibitem [{\citenamefont {Hern{\'{a}}ndez-Garc{\'{i}}a}\ \emph
  {et~al.}(2010)\citenamefont {Hern{\'{a}}ndez-Garc{\'{i}}a}, \citenamefont
  {P{\'{e}}rez-Hern{\'{a}}ndez}, \citenamefont {Ramos}, \citenamefont {Jarque},
  \citenamefont {Roso},\ and\ \citenamefont {Plaja}}]{Hernandez-Garcia2010}%
  \BibitemOpen
  \bibfield  {author} {\bibinfo {author} {\bibfnamefont {C.}~\bibnamefont
  {Hern{\'{a}}ndez-Garc{\'{i}}a}}, \bibinfo {author} {\bibfnamefont {J.~A.}\
  \bibnamefont {P{\'{e}}rez-Hern{\'{a}}ndez}}, \bibinfo {author} {\bibfnamefont
  {J.}~\bibnamefont {Ramos}}, \bibinfo {author} {\bibfnamefont {E.~C.}\
  \bibnamefont {Jarque}}, \bibinfo {author} {\bibfnamefont {L.}~\bibnamefont
  {Roso}}, \ and\ \bibinfo {author} {\bibfnamefont {L.}~\bibnamefont {Plaja}},\
  }\href {\doibase 10.1103/PhysRevA.82.033432} {\bibfield  {journal} {\bibinfo
  {journal} {Physical Review A}\ }\textbf {\bibinfo {volume} {82}},\ \bibinfo
  {pages} {1} (\bibinfo {year} {2010})}\BibitemShut {NoStop}%
\bibitem [{\citenamefont {Hern\'{a}ndez-Garc\'{i}a}\ \emph
  {et~al.}(2017)\citenamefont {Hern\'{a}ndez-Garc\'{i}a}, \citenamefont
  {Turpin}, \citenamefont {Rom\'{a}n}, \citenamefont {Pic\'{o}n}, \citenamefont
  {Drevinskas}, \citenamefont {Cerkauskaite}, \citenamefont {Kazansky},
  \citenamefont {Durfee},\ and\ \citenamefont {{n}igo
  J.~Sola}}]{HernandezGarcia2017}%
  \BibitemOpen
  \bibfield  {author} {\bibinfo {author} {\bibfnamefont {C.}~\bibnamefont
  {Hern\'{a}ndez-Garc\'{i}a}}, \bibinfo {author} {\bibfnamefont
  {A.}~\bibnamefont {Turpin}}, \bibinfo {author} {\bibfnamefont {J.~S.}\
  \bibnamefont {Rom\'{a}n}}, \bibinfo {author} {\bibfnamefont {A.}~\bibnamefont
  {Pic\'{o}n}}, \bibinfo {author} {\bibfnamefont {R.}~\bibnamefont
  {Drevinskas}}, \bibinfo {author} {\bibfnamefont {A.}~\bibnamefont
  {Cerkauskaite}}, \bibinfo {author} {\bibfnamefont {P.~G.}\ \bibnamefont
  {Kazansky}}, \bibinfo {author} {\bibfnamefont {C.~G.}\ \bibnamefont
  {Durfee}}, \ and\ \bibinfo {author} {\bibfnamefont {I.}~\bibnamefont {{n}igo
  J.~Sola}},\ }\href {\doibase 10.1364/OPTICA.4.000520} {\bibfield  {journal}
  {\bibinfo  {journal} {Optica}\ }\textbf {\bibinfo {volume} {4}},\ \bibinfo
  {pages} {520} (\bibinfo {year} {2017})}\BibitemShut {NoStop}%
\bibitem [{\citenamefont {Rego}\ \emph {et~al.}(2019)\citenamefont {Rego},
  \citenamefont {Dorney}, \citenamefont {Brooks}, \citenamefont {Nguyen},
  \citenamefont {Liao}, \citenamefont {San~Rom{\'a}n}, \citenamefont {Couch},
  \citenamefont {Liu}, \citenamefont {Pisanty}, \citenamefont {Lewenstein},
  \citenamefont {Plaja}, \citenamefont {Kapteyn}, \citenamefont {Murnane},\
  and\ \citenamefont {Hern{\'a}ndez-Garc{\'\i}a}}]{Rego2019torque}%
  \BibitemOpen
  \bibfield  {author} {\bibinfo {author} {\bibfnamefont {L.}~\bibnamefont
  {Rego}}, \bibinfo {author} {\bibfnamefont {K.~M.}\ \bibnamefont {Dorney}},
  \bibinfo {author} {\bibfnamefont {N.~J.}\ \bibnamefont {Brooks}}, \bibinfo
  {author} {\bibfnamefont {Q.~L.}\ \bibnamefont {Nguyen}}, \bibinfo {author}
  {\bibfnamefont {C.-T.}\ \bibnamefont {Liao}}, \bibinfo {author}
  {\bibfnamefont {J.}~\bibnamefont {San~Rom{\'a}n}}, \bibinfo {author}
  {\bibfnamefont {D.~E.}\ \bibnamefont {Couch}}, \bibinfo {author}
  {\bibfnamefont {A.}~\bibnamefont {Liu}}, \bibinfo {author} {\bibfnamefont
  {E.}~\bibnamefont {Pisanty}}, \bibinfo {author} {\bibfnamefont
  {M.}~\bibnamefont {Lewenstein}}, \bibinfo {author} {\bibfnamefont
  {L.}~\bibnamefont {Plaja}}, \bibinfo {author} {\bibfnamefont {H.~C.}\
  \bibnamefont {Kapteyn}}, \bibinfo {author} {\bibfnamefont {M.~M.}\
  \bibnamefont {Murnane}}, \ and\ \bibinfo {author} {\bibfnamefont
  {C.}~\bibnamefont {Hern{\'a}ndez-Garc{\'\i}a}},\ }\href {\doibase
  10.1126/science.aaw9486} {\bibfield  {journal} {\bibinfo  {journal}
  {Science}\ }\textbf {\bibinfo {volume} {364}},\ \bibinfo {pages} {eaaw9486}
  (\bibinfo {year} {2019})}\BibitemShut {NoStop}%
\bibitem [{\citenamefont {L'Huillier}\ \emph {et~al.}(1992)\citenamefont
  {L'Huillier}, \citenamefont {Balcou}, \citenamefont {Candel}, \citenamefont
  {Schafer},\ and\ \citenamefont {Kulander}}]{Huillier1992}%
  \BibitemOpen
  \bibfield  {author} {\bibinfo {author} {\bibfnamefont {A.}~\bibnamefont
  {L'Huillier}}, \bibinfo {author} {\bibfnamefont {P.}~\bibnamefont {Balcou}},
  \bibinfo {author} {\bibfnamefont {S.}~\bibnamefont {Candel}}, \bibinfo
  {author} {\bibfnamefont {K.~J.}\ \bibnamefont {Schafer}}, \ and\ \bibinfo
  {author} {\bibfnamefont {K.~C.}\ \bibnamefont {Kulander}},\ }\href {\doibase
  10.1103/PhysRevA.46.2778} {\bibfield  {journal} {\bibinfo  {journal} {Phys.
  Rev. A}\ }\textbf {\bibinfo {volume} {46}},\ \bibinfo {pages} {2778}
  (\bibinfo {year} {1992})}\BibitemShut {NoStop}%
\bibitem [{\citenamefont {Lewenstein}\ \emph {et~al.}(1995)\citenamefont
  {Lewenstein}, \citenamefont {Salieres},\ and\ \citenamefont
  {L'huillier}}]{lewenstein1995phase}%
  \BibitemOpen
  \bibfield  {author} {\bibinfo {author} {\bibfnamefont {M.}~\bibnamefont
  {Lewenstein}}, \bibinfo {author} {\bibfnamefont {P.}~\bibnamefont
  {Salieres}}, \ and\ \bibinfo {author} {\bibfnamefont {A.}~\bibnamefont
  {L'huillier}},\ }\href@noop {} {\bibfield  {journal} {\bibinfo  {journal}
  {Physical Review A}\ }\textbf {\bibinfo {volume} {52}},\ \bibinfo {pages}
  {4747} (\bibinfo {year} {1995})}\BibitemShut {NoStop}%
\bibitem [{\citenamefont {Porras}\ and\ \citenamefont
  {Jolly}(2024)}]{Porras_PRA_2024}%
  \BibitemOpen
  \bibfield  {author} {\bibinfo {author} {\bibfnamefont {M.~A.}\ \bibnamefont
  {Porras}}\ and\ \bibinfo {author} {\bibfnamefont {S.~W.}\ \bibnamefont
  {Jolly}},\ }\href {\doibase 10.1103/PhysRevA.109.033514} {\bibfield
  {journal} {\bibinfo  {journal} {Phys. Rev. A}\ }\textbf {\bibinfo {volume}
  {109}},\ \bibinfo {pages} {033514} (\bibinfo {year} {2024})}\BibitemShut
  {NoStop}%
\bibitem [{\citenamefont {Bliokh}(2025)}]{BliokhPhysLettsA}%
  \BibitemOpen
  \bibfield  {author} {\bibinfo {author} {\bibfnamefont {K.}~\bibnamefont
  {Bliokh}},\ }\href@noop {} {\bibfield  {journal} {\bibinfo  {journal}
  {Physics Letters A}\ }\textbf {\bibinfo {volume} {542}},\ \bibinfo {pages}
  {130425} (\bibinfo {year} {2025})}\BibitemShut {NoStop}%
\end{thebibliography}%


\begin{thebibliography}{9}%
\makeatletter
\providecommand \@ifxundefined [1]{%
 \@ifx{#1\undefined}
}%
\providecommand \@ifnum [1]{%
 \ifnum #1\expandafter \@firstoftwo
 \else \expandafter \@secondoftwo
 \fi
}%
\providecommand \@ifx [1]{%
 \ifx #1\expandafter \@firstoftwo
 \else \expandafter \@secondoftwo
 \fi
}%
\providecommand \natexlab [1]{#1}%
\providecommand \enquote  [1]{``#1''}%
\providecommand \bibnamefont  [1]{#1}%
\providecommand \bibfnamefont [1]{#1}%
\providecommand \citenamefont [1]{#1}%
\providecommand \href@noop [0]{\@secondoftwo}%
\providecommand \href [0]{\begingroup \@sanitize@url \@href}%
\providecommand \@href[1]{\@@startlink{#1}\@@href}%
\providecommand \@@href[1]{\endgroup#1\@@endlink}%
\providecommand \@sanitize@url [0]{\catcode `\\12\catcode `\$12\catcode
  `\&12\catcode `\#12\catcode `\^12\catcode `\_12\catcode `\%12\relax}%
\providecommand \@@startlink[1]{}%
\providecommand \@@endlink[0]{}%
\providecommand \url  [0]{\begingroup\@sanitize@url \@url }%
\providecommand \@url [1]{\endgroup\@href {#1}{\urlprefix }}%
\providecommand \urlprefix  [0]{URL }%
\providecommand \Eprint [0]{\href }%
\providecommand \doibase [0]{http://dx.doi.org/}%
\providecommand \selectlanguage [0]{\@gobble}%
\providecommand \bibinfo  [0]{\@secondoftwo}%
\providecommand \bibfield  [0]{\@secondoftwo}%
\providecommand \translation [1]{[#1]}%
\providecommand \BibitemOpen [0]{}%
\providecommand \bibitemStop [0]{}%
\providecommand \bibitemNoStop [0]{.\EOS\space}%
\providecommand \EOS [0]{\spacefactor3000\relax}%
\providecommand \BibitemShut  [1]{\csname bibitem#1\endcsname}%
\let\auto@bib@innerbib\@empty
\bibitem [{\citenamefont {Porras}(2023)}]{porras2023transverse}%
  \BibitemOpen
  \bibfield  {author} {\bibinfo {author} {\bibfnamefont {M.~A.}\ \bibnamefont
  {Porras}},\ }\href@noop {} {\bibfield  {journal} {\bibinfo  {journal} {Prog.
  Electromagn. Res.}\ }\textbf {\bibinfo {volume} {177}},\ \bibinfo {pages}
  {95} (\bibinfo {year} {2023})}\BibitemShut {NoStop}%
\bibitem [{\citenamefont {Porras}(2024)}]{porras2024clarification}%
  \BibitemOpen
  \bibfield  {author} {\bibinfo {author} {\bibfnamefont {M.}~\bibnamefont
  {Porras}},\ }\href@noop {} {\bibfield  {journal} {\bibinfo  {journal}
  {Journal of Optics}\ }\textbf {\bibinfo {volume} {26}},\ \bibinfo {pages}
  {095601} (\bibinfo {year} {2024})}\BibitemShut {NoStop}%
\bibitem [{\citenamefont {Bliokh}(2012)}]{Bliokh2012}%
  \BibitemOpen
  \bibfield  {author} {\bibinfo {author} {\bibfnamefont {K.}~\bibnamefont
  {Bliokh}},\ }\href@noop {} {\bibfield  {journal} {\bibinfo  {journal}
  {Physical Review A}\ }\textbf {\bibinfo {volume} {86}},\ \bibinfo {pages}
  {033824} (\bibinfo {year} {2012})}\BibitemShut {NoStop}%
\bibitem [{\citenamefont {Bliokh}(2021)}]{Bliokh2021}%
  \BibitemOpen
  \bibfield  {author} {\bibinfo {author} {\bibfnamefont {K.}~\bibnamefont
  {Bliokh}},\ }\href@noop {} {\bibfield  {journal} {\bibinfo  {journal}
  {Physical Review Letters}\ }\textbf {\bibinfo {volume} {126}},\ \bibinfo
  {pages} {243601} (\bibinfo {year} {2021})}\BibitemShut {NoStop}%
\bibitem [{\citenamefont {Bliokh}(2023)}]{Bliokh2023}%
  \BibitemOpen
  \bibfield  {author} {\bibinfo {author} {\bibfnamefont {K.}~\bibnamefont
  {Bliokh}},\ }\href@noop {} {\bibfield  {journal} {\bibinfo  {journal}
  {Physical Review A}\ }\textbf {\bibinfo {volume} {107}},\ \bibinfo {pages}
  {L031501} (\bibinfo {year} {2023})}\BibitemShut {NoStop}%
\bibitem [{\citenamefont {Bliokh}(2025)}]{BliokhPhysLettsA}%
  \BibitemOpen
  \bibfield  {author} {\bibinfo {author} {\bibfnamefont {K.}~\bibnamefont
  {Bliokh}},\ }\href@noop {} {\bibfield  {journal} {\bibinfo  {journal}
  {Physics Letters A}\ }\textbf {\bibinfo {volume} {542}},\ \bibinfo {pages}
  {130425} (\bibinfo {year} {2025})}\BibitemShut {NoStop}%
\bibitem [{\citenamefont {Christov}\ \emph {et~al.}(2001)\citenamefont
  {Christov}, \citenamefont {Bartels}, \citenamefont {Kapteyn},\ and\
  \citenamefont {Murnane}}]{christov2001attosecond}%
  \BibitemOpen
  \bibfield  {author} {\bibinfo {author} {\bibfnamefont {I.}~\bibnamefont
  {Christov}}, \bibinfo {author} {\bibfnamefont {R.}~\bibnamefont {Bartels}},
  \bibinfo {author} {\bibfnamefont {H.}~\bibnamefont {Kapteyn}}, \ and\
  \bibinfo {author} {\bibfnamefont {M.}~\bibnamefont {Murnane}},\ }\href@noop
  {} {\bibfield  {journal} {\bibinfo  {journal} {Physical Review Letters}\
  }\textbf {\bibinfo {volume} {86}},\ \bibinfo {pages} {5458} (\bibinfo {year}
  {2001})}\BibitemShut {NoStop}%
\bibitem [{\citenamefont {Wikmark}\ \emph {et~al.}(2019)\citenamefont
  {Wikmark}, \citenamefont {Guo}, \citenamefont {Vogelsang}, \citenamefont
  {Smorenburg}, \citenamefont {Coudert-Alteirac}, \citenamefont {Lahl},
  \citenamefont {Peschel}, \citenamefont {Rudawski}, \citenamefont {Dacasa},
  \citenamefont {Carlstr{\"o}m} \emph {et~al.}}]{wikmark2019spatiotemporal}%
  \BibitemOpen
  \bibfield  {author} {\bibinfo {author} {\bibfnamefont {H.}~\bibnamefont
  {Wikmark}}, \bibinfo {author} {\bibfnamefont {C.}~\bibnamefont {Guo}},
  \bibinfo {author} {\bibfnamefont {J.}~\bibnamefont {Vogelsang}}, \bibinfo
  {author} {\bibfnamefont {P.~W.}\ \bibnamefont {Smorenburg}}, \bibinfo
  {author} {\bibfnamefont {H.}~\bibnamefont {Coudert-Alteirac}}, \bibinfo
  {author} {\bibfnamefont {J.}~\bibnamefont {Lahl}}, \bibinfo {author}
  {\bibfnamefont {J.}~\bibnamefont {Peschel}}, \bibinfo {author} {\bibfnamefont
  {P.}~\bibnamefont {Rudawski}}, \bibinfo {author} {\bibfnamefont
  {H.}~\bibnamefont {Dacasa}}, \bibinfo {author} {\bibfnamefont
  {S.}~\bibnamefont {Carlstr{\"o}m}},  \emph {et~al.},\ }\href@noop {}
  {\bibfield  {journal} {\bibinfo  {journal} {Proceedings of the National
  Academy of Sciences}\ }\textbf {\bibinfo {volume} {116}},\ \bibinfo {pages}
  {4779} (\bibinfo {year} {2019})}\BibitemShut {NoStop}%
\bibitem [{\citenamefont {Guo}\ \emph {et~al.}(2018)\citenamefont {Guo},
  \citenamefont {Harth}, \citenamefont {Carlström}, \citenamefont {Cheng},
  \citenamefont {Mikaelsson}, \citenamefont {Mårsell}, \citenamefont {Heyl},
  \citenamefont {Miranda}, \citenamefont {Gisselbrecht}, \citenamefont
  {Gaarde}, \citenamefont {Schafer}, \citenamefont {Mikkelsen}, \citenamefont
  {Mauritsson}, \citenamefont {Arnold},\ and\ \citenamefont
  {L’Huillier}}]{Guo_2018}%
  \BibitemOpen
  \bibfield  {author} {\bibinfo {author} {\bibfnamefont {C.}~\bibnamefont
  {Guo}}, \bibinfo {author} {\bibfnamefont {A.}~\bibnamefont {Harth}}, \bibinfo
  {author} {\bibfnamefont {S.}~\bibnamefont {Carlström}}, \bibinfo {author}
  {\bibfnamefont {Y.-C.}\ \bibnamefont {Cheng}}, \bibinfo {author}
  {\bibfnamefont {S.}~\bibnamefont {Mikaelsson}}, \bibinfo {author}
  {\bibfnamefont {E.}~\bibnamefont {Mårsell}}, \bibinfo {author}
  {\bibfnamefont {C.}~\bibnamefont {Heyl}}, \bibinfo {author} {\bibfnamefont
  {M.}~\bibnamefont {Miranda}}, \bibinfo {author} {\bibfnamefont
  {M.}~\bibnamefont {Gisselbrecht}}, \bibinfo {author} {\bibfnamefont {M.~B.}\
  \bibnamefont {Gaarde}}, \bibinfo {author} {\bibfnamefont {K.~J.}\
  \bibnamefont {Schafer}}, \bibinfo {author} {\bibfnamefont {A.}~\bibnamefont
  {Mikkelsen}}, \bibinfo {author} {\bibfnamefont {J.}~\bibnamefont
  {Mauritsson}}, \bibinfo {author} {\bibfnamefont {C.~L.}\ \bibnamefont
  {Arnold}}, \ and\ \bibinfo {author} {\bibfnamefont {A.}~\bibnamefont
  {L’Huillier}},\ }\href {\doibase 10.1088/1361-6455/aa9953} {\bibfield
  {journal} {\bibinfo  {journal} {Journal of Physics B: Atomic, Molecular and
  Optical Physics}\ }\textbf {\bibinfo {volume} {51}},\ \bibinfo {pages}
  {034006} (\bibinfo {year} {2018})}\BibitemShut {NoStop}%
\end{thebibliography}%

%
%
%
%
\end{document}


\title{Supplemental material for Isolated attosecond spatio-temporal optical vortices: interplay between the topological charge and orbital angular momentum scaling in high harmonic generation}

\author{Rodrigo Mart\'{\i}n-Hernández$^{1,2}$, Luis Plaja$^{1,2}$, Carlos Hernández-Garc\'{\i}a$^{1,2}$, Miguel A. Porras$^{3}$}
\address{$^{1}$ Grupo de Investigación en Aplicaciones del Láser y Fotónica, Departamento de F\'{i}sica Aplicada, Universidad de Salamanca, 37008, Salamanca, Spain} 
\address{$^{2}$ Unidad de Excelencia en Luz y Materia Estructuradas (LUMES), Universidad de Salamanca, Salamanca, Spain}
\address{$^3$ Grupo de Sistemas Complejos, ETSIME, Universidad Politécnica de Madrid, Rios Rosas 21, 28003 Madrid, Spain}

\begin{abstract}
    In this Supplementary Material: (1) we provide the analytical expressions used to evaluate the longitudinal and transverse OAM of the the high-order harmonics in Fig. 3 of the main text; (2) we show the harmonic intrisinc t-OAM scaling using the photon centroid, instead of the energy centroind used in the main text, and (3) we show the role of the harmonic intrinsic dipole phase in the generation of the SSOV-driven harmonics in the far-field.
\end{abstract}
\maketitle
\filbreak
\section{1. Orbital angular momentum formulas}

Here we summarize and adapt analytical expressions for the OAM of a field provided in \cite{porras2023transverse,porras2024clarification} that are needed for the evaluation of the OAM of the driver and harmonic fields in HHG.  

For a linearly polarized, paraxial and quasi-monochromatic wave packet of envelope $\psi({\bf x}_\perp, t')$ and carrier angular frequency $\omega_0$ specified in at a transversal plane ${\bf x}_\perp=(x,y)$ at $z$, and local time $t'$, the energy can be evaluated from
\begin{equation}
W= \frac{\varepsilon_0 c}{2} \int |\psi|^2 d{\bf x}_\perp dt',
\end{equation}
where the integral extend to the whole transversal plane and all times. For quasi-monochromatic light, the average number of photons can be approached by $N=W/\hbar \omega_0$.

\subsection{1.1. Longitudinal OAM}

The l-OAM about the a $z$ axis passing through ${\bf x}_{\perp,0}=(x_0,y_0)$ is evaluated from
\begin{equation}
   L_z= \int [(x-x_0)p_y - (y-y_0)p_x] d {\bf x}_\perp dt' = \frac{\varepsilon_0}{2k_0}\int {\rm Im }\left\{\psi^\star[(x-x_0)\partial_y\psi-(y-y_0)\partial_x\psi]\right\} d{\bf x}_\perp dt'
\end{equation}
where ${\bf p}_\perp=(p_x,p_y)$ are the transversal momentum flux densities, and $k_0=\omega_0/c$. The integrals with ${\rm Im}\{\psi^\star \partial_x\psi\}$ are often replaced with integrals with $-i\psi^\star \partial_x\psi$ (and the same for $y$ and $t'$ below), but the later may produce spurious imaginary parts in numerical evaluation. If the wave packet travels along the $z$ direction without misalignment, as in our simulations, the transversal momentum
\begin{equation}
    {\bf P}_\perp = \int {\bf p}_\perp d{\bf x}_\perp dt' = \frac{\varepsilon_0}{2k_0}\int {\rm Im}\{\psi^\star \partial_{{\bf x}_\perp}\psi\} d{\bf x}_\perp dt'
\end{equation}
vanishes, and hence
\begin{equation}
   L_z= \frac{\varepsilon_0}{2k_0}\int {\rm Im }\left\{\psi^\star[x\partial_y\psi-y\partial_x\psi]\right\} d{\bf x}_\perp dt' = L_z^{(i)}
\end{equation}
is independent of the particular $z$ axis, and therefore purely intrinsic.

The average l-OAM and intrinsic l-OAM per photon are then
\begin{equation}
    L_{z,\rm ph}^{(i)} =\frac{L_z^{(i)}}{N}= \frac{\int {\rm Im }\left\{\psi^\star[x\partial_y\psi-y\partial_x\psi]\right\} d{\bf x}_\perp dt'}{\int |\psi|^2 d{\bf x}_\perp dt'},
\end{equation}
in units of $\hbar$.

\subsection{1.2. Transverse OAM}

The transverse OAM about a $y$ axis passing through the origin $(x,z)=(0,0)$, is given by 
\begin{equation}
    L_y = \int (zp_x- x p_z) d{\bf x}_\perp dt' = z P_x -\frac{\varepsilon_0}{2}\int |\psi|^2 x d{\bf x}_\perp dt'=-\frac{\varepsilon_0}{2}\int |\psi|^2 x d{\bf x}_\perp dt',
\end{equation}
since $P_x=0$, and the average l-OAM per photon yields
\begin{equation}
    L_{y,\rm ph} = -k_0 \frac{\int |\psi|^2 x d{\bf x}_\perp dt'}{\int |\psi|^2  d{\bf x}_\perp dt'},
\end{equation}
also in units of $\hbar$.

We can extract the extract the intrinsic, origin independent t-OAM by substracting the extrinsic t-OAM, or t-OAM of the wave spatiotemporal center, say $({\tt x}_{\rm C}, {\tt t}_{\rm C})$, with respect to the $y$ axis $(x,z)=(0,0)$. For a STOV or SSOV traveling at $c$, the extrinsic t-OAM is
\begin{equation}\label{eq:ex}
    L_y^{(e)} = \int [z+c(t'-{\tt t}_{\rm C})]p_xd{\bf x}_\perp dt' - {\tt x}_{\rm C}\int p_zd{\bf x}_\perp dt' =\frac{\epsilon_0}{2k_0}\int {\rm Im}\{\psi^\star ct'\partial_x\psi\}d{\bf x}_\perp dt' - {\tt x}_{\rm C}\frac{\varepsilon_0}{2}\int |\psi|^2 d{\bf x}_\perp dt',
\end{equation}
where we have taken into account that $P_x=0$. 

 Of course, a physically meaningful definition of STOV or SSOV center is necessary. Two alternative choices have been proposed in the literature for ${\tt x}_{\rm C}$ , the energy centroid \cite{porras2023transverse,porras2024clarification} and the photon centroid \cite{Bliokh2012,Bliokh2021,Bliokh2023}, yielding different values of the intrinsic t-OAM for the same STOV/SSOV.

\subsubsection{\bf (a) Energy centroid}

With the choice of the energy centroid,
\begin{equation}
{\tt x}_{\rm EC} = \frac{\int |\psi|^2 x d{\bf x}_\perp dt'}{\int |\psi|^2 d{\bf x}_\perp dt'},
\end{equation}
the extrinsic t-OAM simplifies to
\begin{equation}
    L_{y,\rm EC}^{(e)}=\frac{\epsilon_0}{2k_0}\int {\rm Im}\{\psi^\star ct'\partial_x\psi\}d{\bf x}_\perp dt' - \frac{\varepsilon_0}{2}\int |\psi|^2 x d{\bf x}_\perp dt'.
\end{equation}
Then the intrinsic t-AOM, $L_{y,\rm EC}^{(i)} = L_y - L_{y,\rm EC}^{(e)}$, is given by
\begin{equation}
    L_{y,\rm EC}^{(i)}= -\frac{\epsilon_0}{2k_0}\int {\rm Im}\{\psi^\star ct'\partial_x\psi\}d{\bf x}_\perp dt',
\end{equation}
or, evaluated per photon, 
\begin{equation}
    L_{y,{\rm EC,ph}}^{(i)} = - \frac{\int {\rm Im}\{\psi^\star ct'\partial_x\psi\}d{\bf x}_\perp dt'}{\int |\psi|^2 d{\bf x}_\perp dt'},
\end{equation}
in units of $\hbar$.

\subsubsection{\bf (b) Photon centroid}

The photon centroid has to be evaluated in reciprocal space, ${\bf k}=(k_x,k_y,k_z)$ with the dispersion relation $\omega=c|{\bf k}|$ for the photon wavefunction $\hat E({\bf k})/\sqrt{\omega}$ \cite{Bliokh2012,Bliokh2021,Bliokh2023,BliokhPhysLettsA}. From a practical perspective, and for a paraxial and quasimonochromatic field, it suffices to replace the temporal Fourier transform $\hat \psi(\Omega)$ of the envelope $\psi(t')$ with $\hat\psi(\Omega)/\sqrt{\omega}$, where $\omega=\omega_0+\Omega$, and $\Omega$ is the detuning from $\omega_0$, to write the photon centroid as
\begin{equation}
    {\tt x}_{\rm PC} = \frac{\int (|\hat\psi|^2/\omega) d{\bf x}_\perp d\Omega}{\int (|\hat\psi|^2/\omega) d{\bf x}_\perp d\Omega} .
\end{equation}
For a quasi-monochromatic field one can further approximate $1/\omega \simeq (1/\omega_0)(1-\Omega/\omega_0)$ in the numerator, and $1/\omega\simeq 1/\omega_0$ in the denominator, to obtain,
\begin{equation}
    {\tt x}_{\rm PC} = {\tt x}_{\rm EC} - \frac{1}{\omega_0}\frac{\int |\hat \psi|^2 x\Omega d{\bf x}_\perp d\Omega}{\int |\hat \psi|^2 d{\bf x}_\perp d\Omega} = {\tt x}_{\rm EC} + \frac{1}{\omega_0}\frac{\int {\rm Im}\{\psi^\star x\partial_{t'}\psi\} d{\bf x}_\perp dt'}{\int |\psi|^2 d{\bf x}_\perp dt'}.
\end{equation}
The last expression provides the location of the photon centroid from the direct spatiotemporal field, yielding expected values such as ${\tt x}_{\rm PC} =0$ for a SSOV without any vortex, and the known displacement $\ell /(2k_0\gamma)$ from ${\tt x}_{\rm EC}$ proportional to the topological charge for a STOV of ellipticity $\gamma=c t_0/x_0$ \cite{Bliokh2023}.

When the generic ${\tt x}_{\rm C}$ is replaced with the photon centroid ${\tt x}_{\rm PC}$, Eq. (\ref{eq:ex}) leads to the extrinsic t-OAM
\begin{equation}
    L_{y,\rm PC}^{(e)}=\frac{\epsilon_0}{2k_0}\int {\rm Im}\{\psi^\star( ct'\partial_x- x\partial_{ct'})\psi\}d{\bf x}_\perp dt' - \frac{\varepsilon_0}{2}\int |\psi|^2 x d{\bf x}_\perp dt',
\end{equation}
and $L_y^{(i)} = L_y- L_{y,\rm PC}^{(e)}$ yields
\begin{equation}
    L_{y,\rm PC}^{(i)} = -\frac{\varepsilon_0}{2k_0}\int {\rm Im}\left\{\psi^\star(ct'\partial_x-x \partial_{ct'})\psi \right\} d{\bf x_\perp}dt', 
\end{equation}
or, expressed per photon and in units of $\hbar$,
\begin{equation}
    \label{eq:Ly_pc}
    L_{y,{\rm PC,ph}}^{(i)} = - \frac{\int {\rm Im}\{\psi^\star (ct' \partial_x- x\partial_{ct'})\psi\}d{\bf x}_\perp dt'}{\int |\psi|^2 d{\bf x}_\perp dt'}.
\end{equation}

\section{2. Photon centroid-based intrinsic \lowercase{t}-OAM scaling and non-conservation in HHG}
\begin{figure*}[b]
\includegraphics[width=0.9\textwidth]{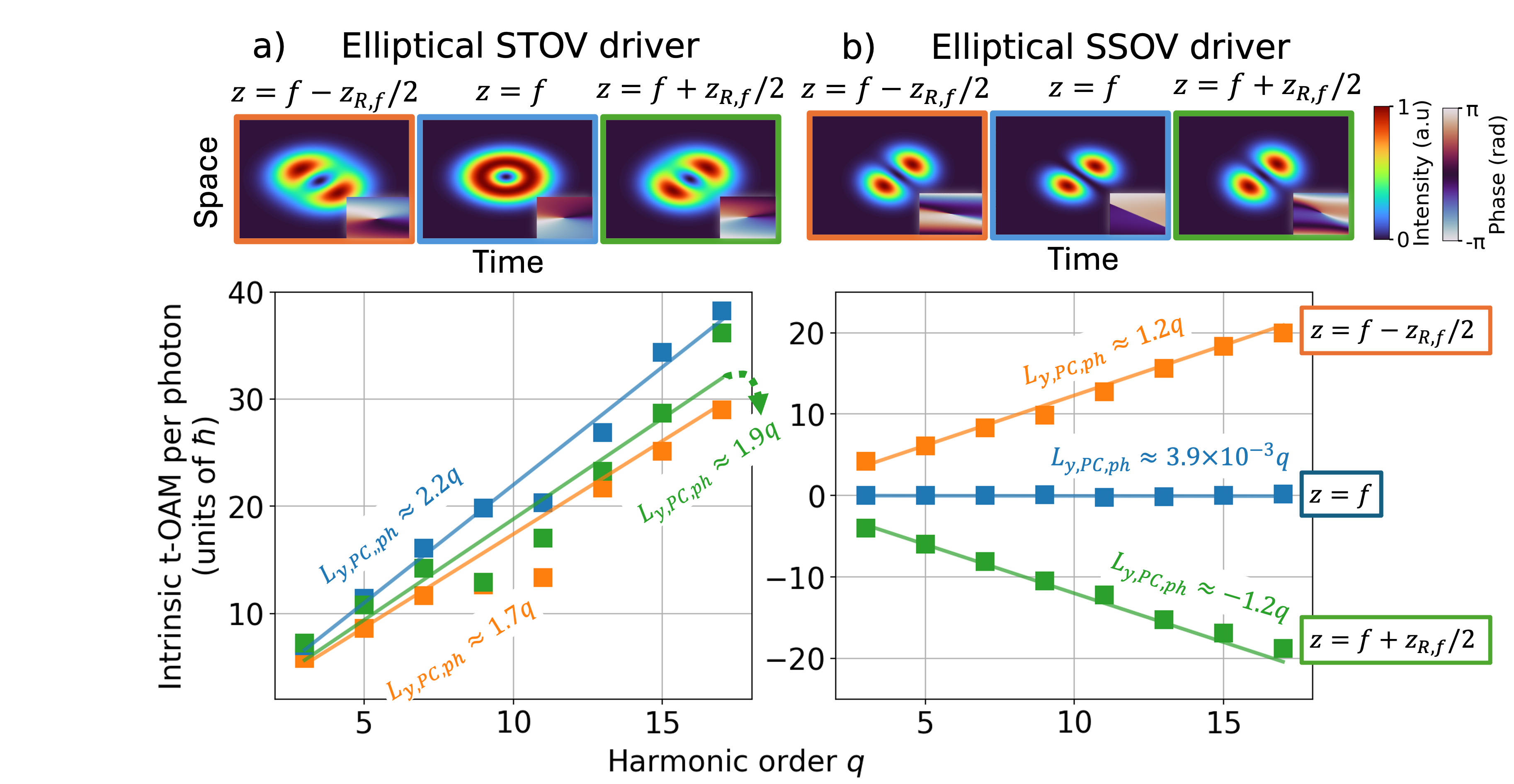}
\caption{Comparative of the harmonic average intrinsic, PC-based, t-OAM per photon scaling for (a) STOV, and (b) SSOV as drivers, with the gas-jet placed at distances $f-z_{R,f}/2$ (orange), $f$ (blue) and $f+z_{R,f}/2$ (green), where $z_{R,f}=f^2/z_R$ is the focal Rayleigh range. The intensity and phase of the respective drivers are shown in the top panels. The solid lines are the predictions of the elemental model and the symbols of the advanced numerical simulations. The driving field parameters are the same as in Fig. 3.} \label{fig1}
\end{figure*}

We have repeated the whole analysis detailed in the main text on the intrinsic t-OAM per photon scaling and intrinsic t-OAM conservation in the same HHG processes with the STOV and SSOV drivers of Fig. 3 (b) and (c), but taking the photon centroid to extract the intrinsic t-OAM, or photon centroid-based (PC-based) intrinsic t-OAM. Equation  (\ref{eq:Ly_pc}) is used to evaluate the PC-based intrinsic t-OAM per photon of the driver and the harmonics just at the three gas jet positions, since the PC-based intrinsic t-OAM is generally not conserved in free space propagation \cite{porras2024clarification,BliokhPhysLettsA} (see also below). The results are shown in Fig. \ref{fig1}. With the same STOV and SSOV, the total t-OAM continues to be zero, and the extrinsic t-OAM is again opposite to the PC-based intrinsic t-OAM; hence only the PC-based intrinsic t-OAM per photon is shown. The PC-based intrinsic t-OAM of the STOV and SSOV drivers is computed to be different at the three gas jet positions, so that the intrinsic t-OAM conservation would require the three different scalings $L_{y,\rm PC,ph}(q)\approx 2.1q$ at $z=f-z_{R,f}/2$, $L_{y,\rm PC,ph}(q)\approx  2.2q$ at $z=f$ and $L_{y,\rm PC,ph}(q)\approx 2.3 q$ at $z=f+z_{R,f}/2$ where the three numbers are the PC-based intrinsic t-OAM per photon ($\hbar$ units) of the STOV driver at the three positions, and $L_{y,\rm PC,ph}(q)\approx 2.1 q$ at $z=f-z_{R,f}/2$, $L_{y,\rm PC,ph}(q)\approx 3.9\times 10^{-3}q$ at $z=f$, and $L_{y,\rm PC,ph}(q)\approx-2.1 q$ at $z=f+z_{R,f}/2$, with the three numbers the same for the SSOV driver. Non-conservation of the PC-based intrinsic t-OAM of the driver particularly enhanced for the SSOVs driver. As with the energy centroid-based, we observe in Figs. \ref{fig1} (a) and (b) a general linear scaling with $q$ with the indicated slopes (color lines and symbols), but the scaled intrinsic t-OAM per photon is not generally that of the drivers, implying again non-conservation in the up-conversion process. Note that the conservation of the intrinsic t-OAM per photon holds only if the generation position is the focus of the driving beam, $z=f$.

\section{3. Role of intrinsic dipole phase}

\begin{figure*}[t]
\includegraphics[width=0.9\textwidth]{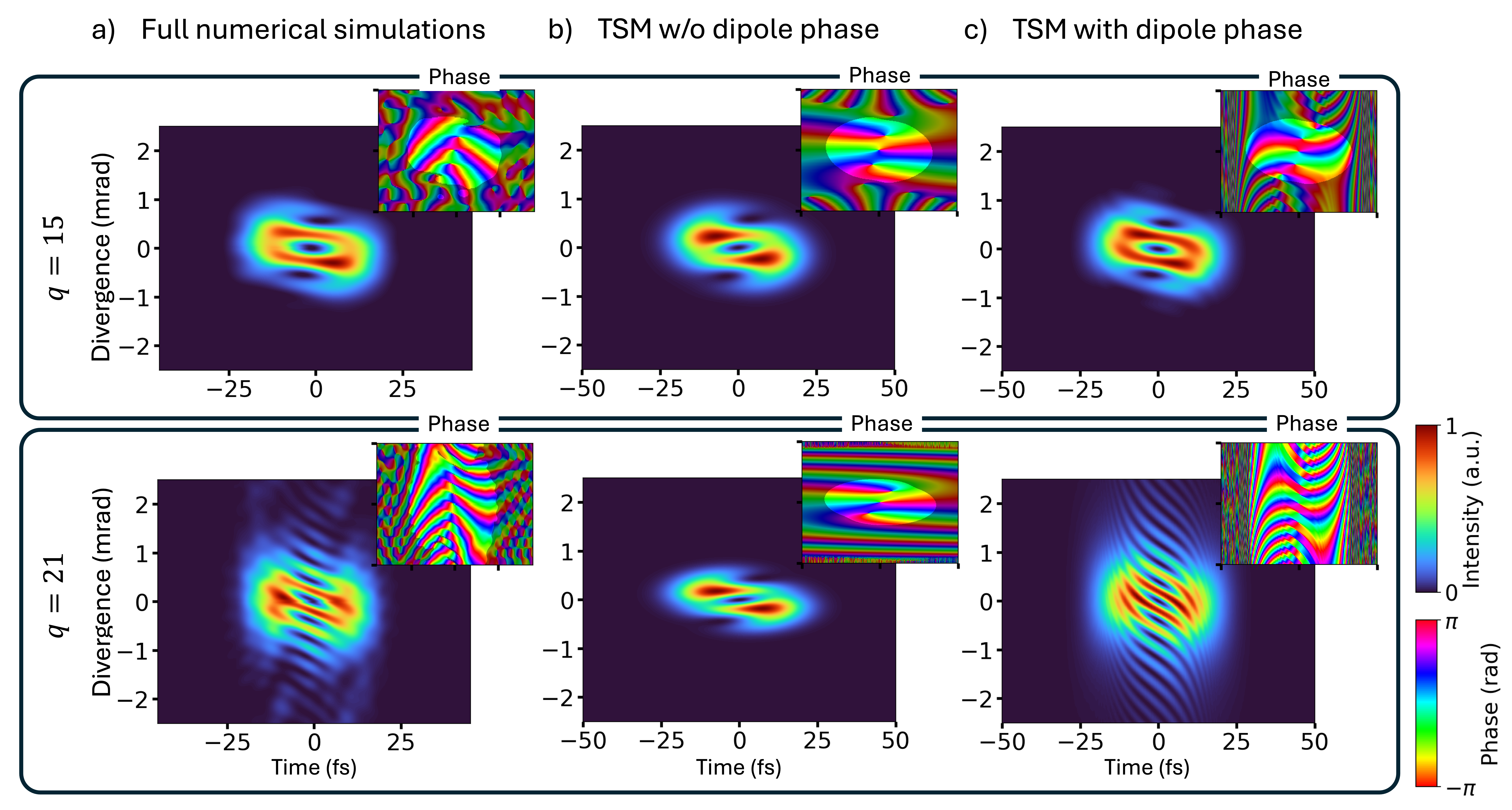}
\caption{Effect of the intrinsic dipole phase into the far-field EUV-STOVs spatiotemporal distribution, calculated using a) the full numerical simulations, b) the TSM without the intrinsic dipole phase and c) the TSM considering the intrinsic dipole phase. We show the comparison for the 15$^{\rm th}$ (top row) and 21$^{\rm st}$ (bottom row) harmonic orders. The parameters of driving SSOV beam ($\eta = 2.0$) are the same as in Fig. 1. The pase insets are masked to highlight the region where the harmonic intensity is above $10^{-2}$ a.u.} \label{fig:dip_phase}
\end{figure*}

The intrinsic dipole phase refers to the phase accumulated by the electron during its excursion in the continuum, which is subsequently imprinted into the emitted harmonic radiation upon recombination step. Although it is fundamentally a microscopic effect, the dipole phase plays a notable role in understanding the macroscopic spatiotemporal characteristics of the HHG emission \cite{christov2001attosecond, wikmark2019spatiotemporal, Guo_2018}. This phase strongly depends on the driver's intensity  and with the type of electron trajectory --- short or long. When Gaussian beam profiles are employed for driving the HHG process,  the spatial intensity distribution of the driving laser field results in a intrinsic dipole phase that introduces a curvature to the harmonic wavefront, effectively acting as a diverging lens. Upon propagation to the far field, this leads to harmonic divergences that generally increase with harmonic order \cite{wikmark2019spatiotemporal}.

In Fig \ref{fig:dip_phase}, we compare the far-field spatiotemporal harmonic distributions from  (a), the full advanced numerical simulations, (b), the TSM without the dipole phase contribution, and (c), taking into consideration the dipole phase  for the 15$^{\rm th}$ and 21$^{\rm st}$ harmonic orders. We employ the $\gamma$ model \cite{wikmark2019spatiotemporal} to estimate the dipole phase for the short trajectories, which are known to be the most relevant in the macroscopic HHG emission. We observe that intrinsic dipole phase, while non-negligible for the 15$^{\rm th}$ harmonic order, its effect becomes more noticeable for harmonics closer to the cut-off ($q_{\rm cutoff} = 27$). The imprinted diverging wavefront populates more the satellite spatiotemporal phase singularities, resulting in a vertically aligned array of harmonic STOVs. Only in the latter case, the TSM with the dipole phase, is able to properly describe the results from the full numerical simulations.

\bibliography{supplemental}